\documentclass[a4paper,fleqn,usenatbib]{mnras}

\usepackage{newtxtext,newtxmath}

\usepackage[T1]{fontenc}
\usepackage{ae,aecompl}

\usepackage{siunitx}
\usepackage{graphicx}	
\usepackage{amsmath}	
\usepackage{amssymb}	

\title[The dynamics of A ring edge perturbed by 
Janus and Epimetheus]{The dynamics of the outer edge of 
Saturn's A ring perturbed by the satellites Janus and Epimetheus}

\author[Araujo et al.]{%
  N. C. S. Araujo$^{1}$;
  S. Renner$^{2}$\thanks{E-mail: Stefan.Renner@univ-lille.fr};
  N. J. Cooper$^{3}$;
  M. El Moutamid$^{4}$; 
  C.D. Murray $^{3}$; 
  \newauthor
  B. Sicardy$^{5}$
  and
  E. Vieira Neto$^{1}$ \\ 
  $^{1}$ Departamento de Matem\'{a}tica, S\~{a}o Paulo State 
  University (UNESP), Campus Guaratinguet\'{a}, 12516-410 
  Guaratinguet\'{a}, S\~{a}o Paulo, Brazil \\
  $^{2}$ IMCCE, Observatoire de Paris, CNRS UMR 8028, Universit\'e de Lille, 
  Observatoire de Lille, 1 impasse de l'Observatoire, 59000 Lille, France\\
  $^{3}$  Astronomy Unit, School of Physics and Astronomy, 
  Queen Mary University of London, Mile End Road, London, E1 4NS, UK \\
  $^{4}$ Center for Astrophysics and Planetary Science,
  Carl Sagan Institute, Cornell University, Ithaca, NY 14853, USA\\
 $^{5}$ LESIA, Observatoire de Paris, Universit\'e PSL, CNRS, Sorbonne Universit\'e, Universit\'e Paris Diderot, Sorbonne Paris Cit\'e, 5 place Jules \\ Janssen, 92195 Meudon, France}
\begin{document}

\date{Accepted 2019 April 24. Received 2019 April 24; in original form 2018 October 11}

\pagerange{\pageref{firstpage}--\pageref{lastpage}} \pubyear{2019}

\maketitle

\label{firstpage}

\begin{abstract}
 We present an analytical model to study the dynamics of the outer
edge of Saturn's A ring. The latter is influenced by 7:6 mean motion
resonances with Janus and Epimetheus. Because of the horseshoe motion
of the two co-orbital moons, the ring edge 
particles are alternately trapped in a corotation eccentricity 
resonance (CER) or a
Lindblad eccentricity resonance (LER). However, the resonance oscillation
periods are longer than the four-year interval
between the switches in the orbits of Janus and Epimetheus. Averaged
equations of motion are used, and our model is numerically integrated
to describe the effects of the periodic sweeping of the 7:6 CERs and
LERs over the ring edge region. We show that four radial zones (ranges
136715-136723, 136738-136749, 136756-136768, 136783-136791 km) are
chaotic on decadal timescales, within which particle semi-major axes
have periodic changes due to partial libration motions around the CER
fixed points. After a few decades, the maximum variation of semi-major
axis is about eleven (resp. three) kilometres in the case of the CER
with Janus (resp. Epimetheus). Similarly, particle eccentricities have
partial oscillations forced by the LERs every four years. 
For initially circular orbits, the
maximum eccentricity reached is $\sim 0.001$. 
We apply our work to ``Peggy", an object recently discovered 
at the ring
edge \citep{Murray.etal-2014}, confirming that it is strongly perturbed 
by the Janus 7:6 LER. The CER has currently no effect on that body, 
nevertheless  the 
fitted semi-major axes are just outside the chaotic zone 
of radial range 136756-136768 km. 

\end{abstract}


\section{Introduction}

\begin{figure}
\centerline{\includegraphics[width=0.99\columnwidth,angle=0]{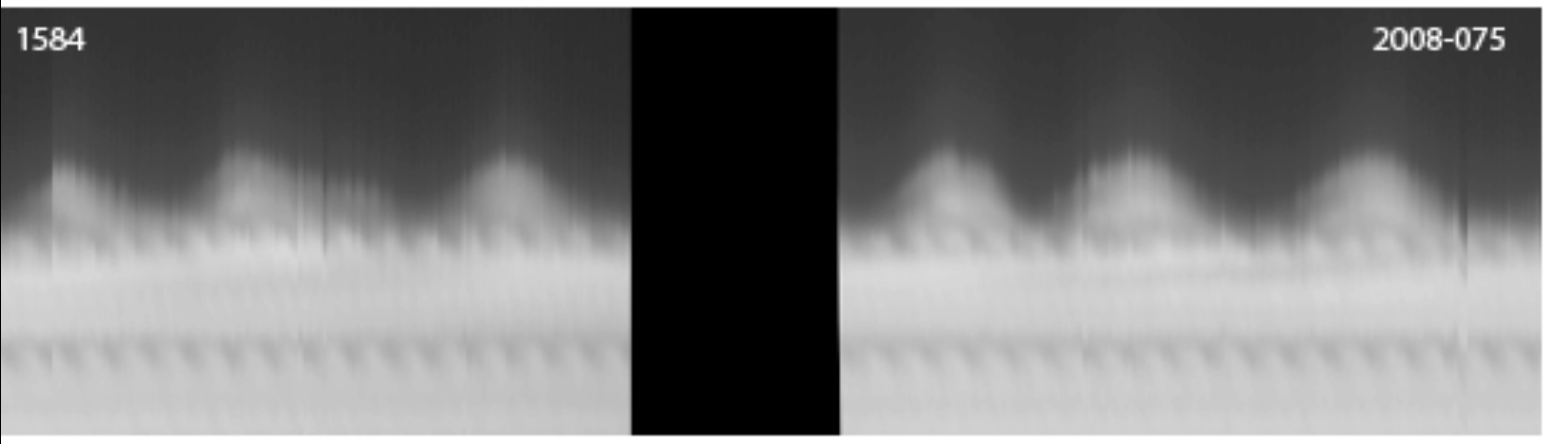}}
\caption{Mosaic of the A ring edge dominated by a 7-lobed pattern, as a result of the 7:6 inner LER with Janus
(note that one lobe is not seen due to a gap in the data). 
The horizontal scale (co-rotating longitude) is from $0$ to $360$ deg and the vertical scale (radial distance) is from $136600$ to $136900$ km.  
The individual images (267 frames from Cassini's Narrow Angle Camera on 2008 March 15) were reprojected and then assembled using a co-rotating frame moving at Janus' mean motion (518.346 deg.day$^{-1}$).
The high frequency black and white patterns are due to the 35:34 and 34:33 inner LERs with Prometheus.}
 \label{lobe7}
\end{figure}

The first resolved images of the inner satellites of Saturn Janus and Epimetheus were provided in 1980 by the Voyager 1 spacecraft
\citep{Smith-1981}.
Small differences in the  mean motions of Janus and Epimetheus (Table \ref{orbitalperiodtable})
cause the two co-orbital satellites to approach one another at intervals of about 4.2 years.
At these times, the mutual perturbations lead to a switch in their orbital configuration.
In fact, in a reference frame centred on Saturn and rotating with the average mean motion, Janus and Epimetheus each follow a horseshoe orbit separated by $180$ degrees in longitude
\citep{Yoder-1983,Murray.Dermott-1999}. 
\citet{Porco.etal-1984} examined the outer edge of Saturn's A ring using the Voyager data.
They showed that this structure is influenced by a 7:6 inner Lindblad eccentricity resonance (LER) with the co-orbital moons 
(Figure \ref{lobe7}), the resonance forcing a seven-lobed radial distortion of amplitude $6.7 \pm 1.5$ km on ring particles that revolve with the average mean motion of the co-orbital satellite system.
\citet{Spitale.Porco-2009} revisited the shape and kinematics
of the ring edge using high-resolution (1-10 km pixel-scale) Cassini ISS (Imaging Science Subsystem) images spanning nearly four years between 2005 and 2009, allowing the variation with time to be measured.
They found that the ring edge is irregular for data obtained within 
about eight months of the encounter between Janus and Epimetheus that occured in 2006, suggesting that a period
of adjustment occurs as the satellites approach and recede from
each switch in their orbits. 
Outside that adjustment period, when Janus is in the inner position of its orbit, the ring is dominated by a $m = 7$ pattern revolving at the same rate as Janus' mean motion.  
However, the alignment is opposite in phase to that predicted for isolated test particles : \citet{Porco.Nicholson-1987} predict that streamlines exterior to the resonance like those at the A ring edge should have one apocentre oriented towards the satellite, instead of one pericentre as observed. This configuration, also seen at the outer edge of the B ring, is most likely the result of ring particle interactions \citep{Porco.etal-1984}.
Finally, the amplitude of the radial distortion varies with time, a behavior that \citet{Spitale.Porco-2009} attributed to 
a beat pattern between the perturbations from the two satellites. 
\citet{ElMoutamid.etal-2016} extended the ring edge analysis using 
Cassini images and occultation data over a period of 8 years from 2006 to 2014. 
Their fits confirmed that, for the period between 2006 and 2010 when Janus is on the inner leg
of the horseshoe orbit, the A ring outer edge is dominated by the Janus 7:6 LER.
However, the 7-lobed pattern disappears after the Janus/Epimetheus orbit swap in 2010 which
moves the Janus resonance away from the ring edge. 
Moreover, \citet{ElMoutamid.etal-2016} found a variety of pattern speeds with different azimuthal
wave numbers, probably representing waves trapped in resonant cavities, and also
identified some other signatures consistent with tesseral resonances that could be associated
with inhomogeneities in Saturn's interior. 

Recently, \cite{Murray.etal-2014} discovered an embedded, sub-km-radius object (nicknamed ``Peggy") at the ring edge in Cassini images producing localized, time-varying structures due to its gravitational perturbation of nearby ring material.
The observational signature of this object has been tracked since its discovery in 2013, 
the deduced semi-major axis varying within 
$\sim 10$ km of the ring edge location, between 136766 km and 136775 km. 
The careful tracking of the features detected leads to longitude patterns that are consistent 
with the orbital evolution of several (at least two) objects. 
Semi-major axis changes could mainly arise from collisions with the local ring material 
due to the differences in relative eccentricity. 
Such collisions could also cause the destruction of an outwardly evolving object due to ring torques. This may explain the unusual brightness of the object seen in the discovery image by \cite{Murray.etal-2014} and the apparent disappearance of another feature soon afterwards. 
Moreover, the periodic sweeping of the Janus/Epimetheus 7:6 resonances may also have an effect over the orbit of the embbeded object.  

This is the core of this paper : predicting the consequences of the 
abrupt satellite semi-major axis changes at four-year intervals over the resonant 
orbits of the A ring edge particles. 
In fact, the strongest resonances from Janus/Epimetheus that affect 
the A ring edge are the Corotation Eccentricity Resonance (hereafter, CER) and the Lindblad Eccentricity Resonance (LER). 
These two types of resonances involve the following arguments :
$\Psi_{\rm C}=$(m+1)$\lambda_{S}$$-$m$\lambda$$-$$\varpi_{S}$ and 
$\Psi_{\rm L}=$(m+1)$\lambda_{S}$$-$m$\lambda$$-$$\varpi$, where
$\lambda$ represents the geometric mean longitude, $\varpi$  the geometric longitude of pericentre, the subscript $S$ stands for the perturbing satellite, and $m=6$ in the case of the ring edge.
The LERs and CERs produce different effects : 
a CER affects the semi-major axis of the particles, forcing 
a libration motion inside the so-called corotation sites, but keeps
orbital eccentricities almost constant. In
contrast, a LER modifies eccentricities and keeps semi-major
axes almost constant \citep{Moutamid.etal-2014}.
 
We focus on providing a short-term dynamical study of the motion of the A ring edge perturbed by the 7:6 CER/LER due to Janus and Epimetheus. Therefore the ring physical effects (collisions, viscosity, self-gravity,..) will not be addressed here. 
In particular, the goal is to understand the additional effect of the periodic trapping into the CERs, which was not considered in previous work. 
The paper is organized as follows. 
Section 2 summarizes the orbital characteristics of Janus-Epimetheus and gives an overview of the CERs/LERs affecting the edge of the A ring. 
Section 3  presents the analytical modeling, basically averaged equations of motion near the resonances and associated timescales.  
Numerical simulation results are examined and the model is applied to object ``Peggy"
in Section 4. 
Concluding remarks are given in Section 5.

\section{Motion of Janus-Epimetheus and resonances at
the A ring edge}

\begin{figure}
\centerline{\includegraphics[width=0.99\columnwidth]{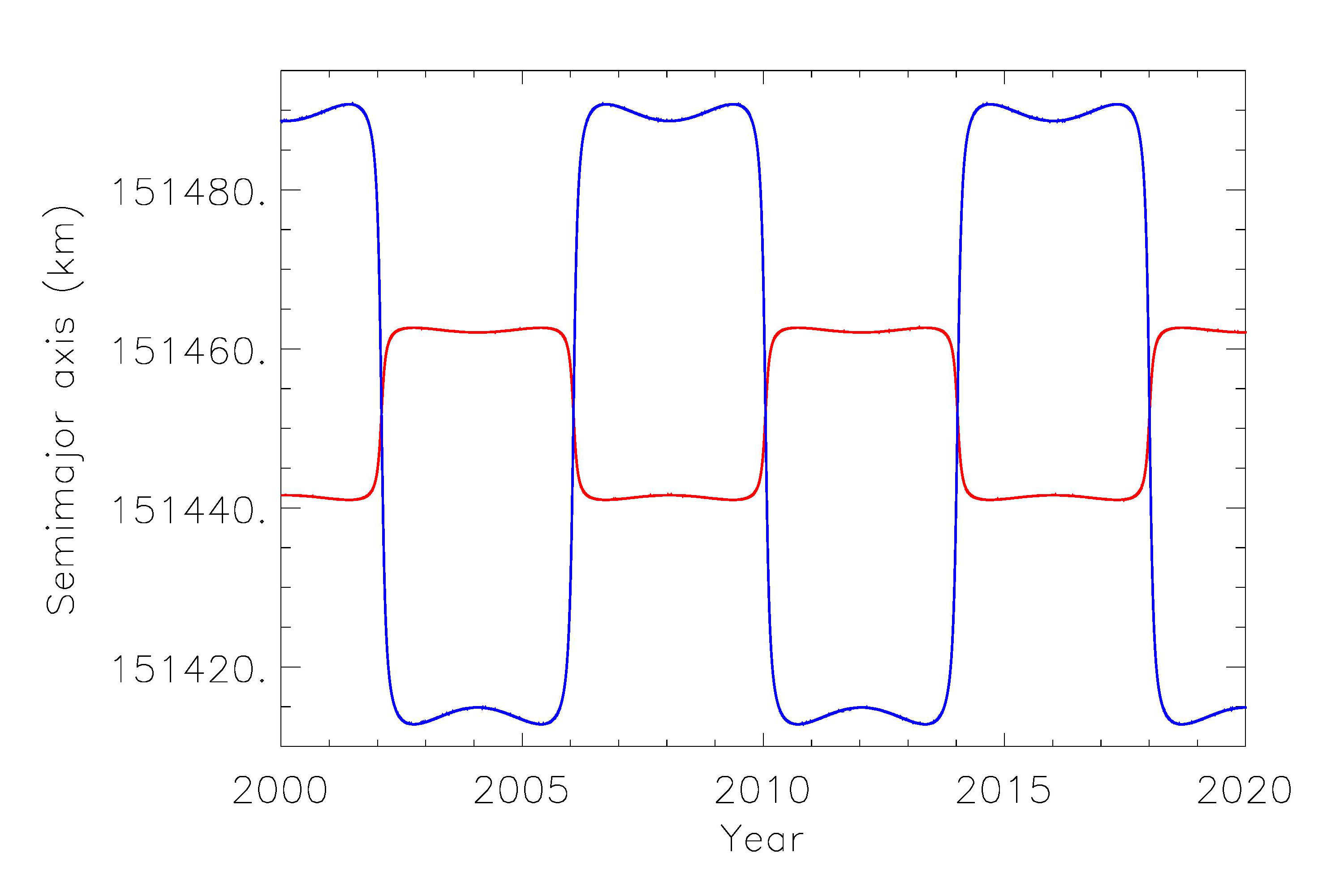}}
\caption{Semi-major axis for Janus (red) and Epimetheus (blue) as a function of time between 2000 and 2020, from a numerical integration fitted to Cassini data \citep{Cooper.etal-2015}. 
The initial conditions and Saturn's parameters are given in Table \ref{tab_orb_elements} and Table \ref{tab_param_saturn}, respectively.}
 \label{jaep_sma}
\end{figure}

Figure \ref{jaep_sma} shows the time evolution of the semi-major axes and the switches in the orbits of Janus and Epimetheus, resulting from a numerical model fitted to Cassini
ISS astrometric data for all the inner satellites \citep{Cooper.etal-2015}.  
For this fit, the state vectors and masses were solved at the
epoch 2007 June 01 00:00:00.0, and ephemerides spanning the
period 2000-2020 were generated. 
As a result, Table \ref{tab_orb_elements} gives the geometric orbital elements 
for Janus and Epimetheus computed from the state vectors using the method of \citet{Renner.Sicardy-2006} and the parameters of
Saturn given in Table \ref{tab_param_saturn}. 

The radial widths of the librational arcs of Janus and Epimetheus in Figure \ref{jaep_sma} have a ratio equal to the mass ratio ($\sim 3.6$, see Table \ref{tab_orb_elements}), as a consequence of the conservation of angular momentum. Thus, Janus has the smaller variations in semi-major axis since it is the more massive moon. 
The difference between the semi-major axes of the two satellites is approximately $48$ km. 
Note that away from the approaches, the radial separation between the satellites is not constant : it is maximum for a relative longitude corresponding to the $L_4$, $L_5$ Lagrangian points, and minimum at the longitude of the $L_3$ point
\citep{Dermott-1981a,Dermott-1981b}. \\

\begin{table} 
\caption{Mean motions and orbital periods for Janus, Epimetheus 
\citep{Cooper.etal-2015} and the outer edge of the A ring \citep{ElMoutamid.etal-2016}.}
\label{orbitalperiodtable}
\begin{center}
\begin{tabular}{l|l|l} 
& Mean motion (deg.day$^{-1}$) & Orbital period  (day) \\
\hline
A ring edge &  604.22 & 0.59581 \\
\hline
Janus &  518.35 or 518.24 & 0.69451 or 0.69466 \\
\hline
Epimetheus & 518.10 or 518.49 & 0.69484 or 0.69432               
\end{tabular}
\end{center}
\end{table}
\begin{table}
\centering
	\caption{Geometric orbital elements for Janus and Epimetheus 
	at epoch 2007 JUN 01 00:00:00.0 UTC (JED 2454252.50075446), 
computed from fits to $Cassini$ observations \citep{Cooper.etal-2015}.
At this epoch, the orbit of Janus is in the inner position.}
\label{tab_orb_elements}	
	\begin{tabular}{l|ll}
 & Janus & Epimetheus  \\
 \hline
Mass (kg) & $1.896 \times 10^{18}$ & $5.262 \times 10^{17}$   \\
\hline
$a$ (km) & 151441.171 & 151488.969   \\
\hline
$e$  & 0.00677 &  0.00972   \\
\hline
$i$ (deg) & 0.16446  &  0.35199   \\
\hline
$\Omega$ (deg) & 16.03786    &  53.92315    \\
\hline
$\varpi$ (deg) & 328.67973 & 79.18426   \\
\hline
$\lambda$ (deg) & 268.22574   &  145.38974      \\
\end{tabular} \\
\raggedright
Note: The elements $a$, $e$, $i$, $\Omega$, $\varpi$, $\lambda$ are 
respectively the 
semi-major axis, the eccentricity, the inclination, the longitude of ascending node, 
the longitude of pericentre, and the mean longitude. 
Mean motion and the pericentre precession rates are derived 
self-consistently from the semi-major axis using the Saturn constants
given in Table \ref{tab_param_saturn}.
\end{table}
\begin{table}
\centering
	\caption{Saturn constants, from \citet{Cooper.etal-2015}.}
	\label{tab_param_saturn}
	\begin{tabular}{l|l|l}
Constant&Value&units  \\
\hline
 $GM$  & $3.793120706585872 \times 10^{7}$  &km$^3$ s$^{-2}$\\
 Radius          &60330  &       km\\
 $J_{2}$         &$1.629084747205768 \times 10^{-2}$ & \\
 $J_{4}$         &$-9.336977208718450 \times 10^{-4}$& \\
 $J_{6}$         &$9.643662444877887\times 10^{-5}$ & \\
\end{tabular}
\end{table}

\begin{figure}
\centerline{\includegraphics[width=0.99\columnwidth,angle=0]{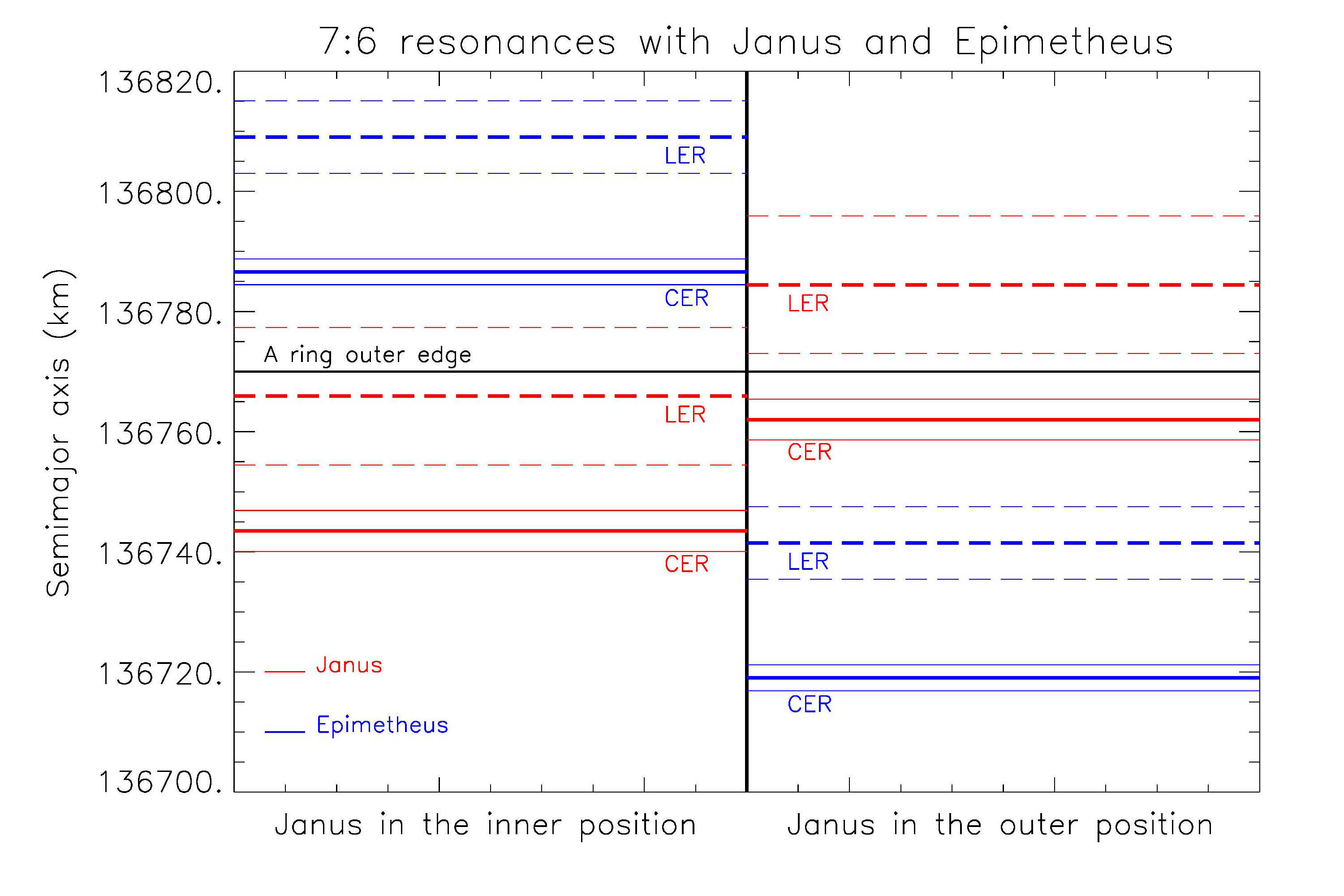}}
\caption{Semi-major axes of the 7:6 resonances with Janus (in red) and Epimetheus (in blue) at the edge of the A ring (solid black line representing its average location, see
\citet{ElMoutamid.etal-2016}). The LERs and CERs appear 
respectively as thick dotted and solid lines, when Janus is in the inner position on the left, or in the outer position on the right. The full widths of the resonances are indicated by thinner lines on either side of the resonance position. See Table \ref{jaep_res_values}	and text for details.}
 \label{jaep_res}
\end{figure}
\begin{table}
\centering
	\caption{Numerical values (km) for the locations of the 7:6 resonances shown Figure \ref{jaep_res}. 
The red (resp. blue) colour applies to Janus (resp. Epimetheus). 
The full widths are in parentheses and are given by Eqs.  (\ref{width_cer}) and
 (\ref{width_ler}) 	. 
The location of the resonances are computed iteratively by cancelling the 
time derivatives of the resonance angles, and using the mean motion values, the pericentre precession rates from Table \ref{orbitalperiodtable} and \ref{tab_orb_elements} and the Saturn
parameters given in Table \ref{tab_param_saturn}. 	
	}
\label{jaep_res_values}	
	\begin{tabular}{l|ll}
 & Janus inside & Janus outside \\
 \hline
LER & {\color{blue} 136809.1 (12.1)} &  {\color{red} 136784.5 (22.9)  } \\
CER & {\color{blue} 136786.6 (4.3)}  &  {\color{red} 136762.0 (6.8)} \\
LER & {\color{red}136765.9 (22.9)} &  {\color{blue} 136741.5 (12.1) }  \\
CER & {\color{red} 136743.5 (6.8)}	  &  {\color{blue} 136719.0 (4.3)  } \\
\end{tabular}
\end{table}

The direct consequence of the horsehoe motion is that the locations of the 7:6 resonances by Janus-Epimetheus shift inwards or outwards every four years. Since the semi-major axis difference between the resonances is comparable to the radial excursions of the two satellites, the particles located in three specific regions (around semi-major axes $\sim 136740$, 136765 and 136785 km) are alternately trapped in a CER or a LER. 
This peculiarity is presented in Figure \ref{jaep_res} and Table \ref{jaep_res_values}, which give the resonance locations and widths. 
For instance, when Janus is in the outer position, the ring edge is roughly at Janus CER position, with the (weaker) Epimetheus LER located inside the ring and the (stronger) Janus one displaced outside, thus explaining the  
smaller eccentricities of the ring edge fitted by \citet{Spitale.Porco-2009} for observations a few months preceding the inward swap of Janus.
As demonstrated in Sections \ref{theory_part} and \ref{num_part}, the CERs also cause significant perturbations on the ring particles, in spite of the low-eccentricity orbits of Janus and Epimetheus. 

\begin{figure}
\includegraphics[width=0.99\columnwidth]{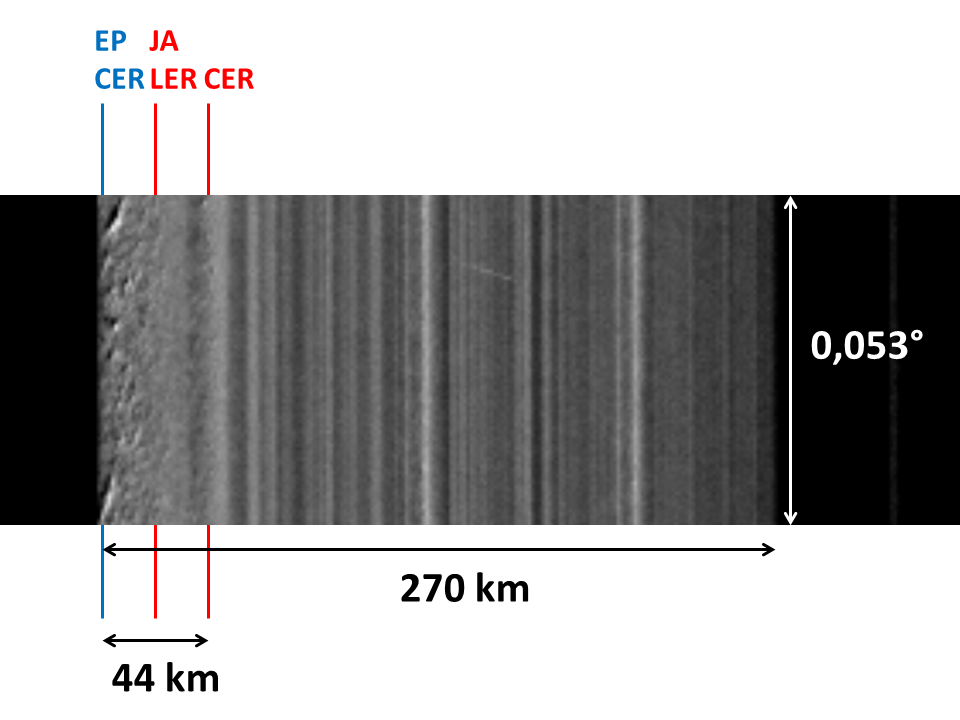}
\includegraphics[width=0.92\columnwidth]{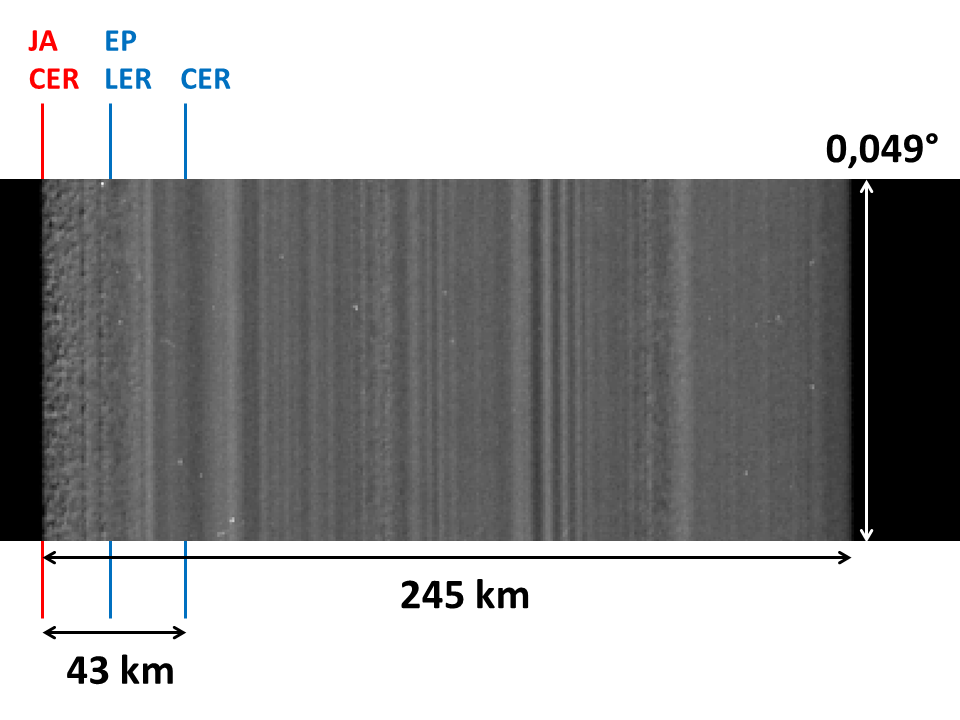}
\caption{Cassini ISS high-resolution images showing narrow parts of the A ring edge. 
The top image was cropped from the Cassini frame
N1591063789\_1, ISS\_070RI\_PAZSCN001\_PRIME, 2008-154T01:30:09.718Z, and the bottom one from 
N1656998640\_1, ISS\_134DA\_DAPHNIS001\_VIMS, 2010-186T04:38:30.079Z. The locations of the 7:6 resonances have been superimposed. 
Janus is in the inner (resp. outer) position at the epoch of the top (resp. bottom) image. 
The black part on the right is the Keeler gap. }
 \label{cassiniobs}
\end{figure}

Figure \ref{cassiniobs} shows two Cassini ISS high-resolution images of the A ring edge
where  the locations of the relevant 7:6 resonances have been superimposed,  one corresponding to Janus in the inner position and the other in the outer position. 
Thanks to a grazing illumination, these data clearly reveal the presence of clumps with vertical extents. 
By selecting another longitude interval to produce the top image, note that the ring edge would instead match the location of the LER with Janus. Indeed the edge is structured by seven lobes of radial amplitude $\sim 15$ km \citep{Spitale.Porco-2009}, a distance comparable to the location difference between the CER with Epimetheus and the LER with Janus. \\
%

We can neglect the inclination of Janus and Epimetheus in this work. 
Indeed, the mean motion inclination resonances are located well inside the edge of the A ring. The strongest of these resonances are respectively the corotation inclination resonance (CIR) and the Lindblad inclination resonance (LIR), for which the critical angles are $\displaystyle \Psi_{\rm CIR}=2[(m+1)\lambda_{\rm S} - m\lambda -\Omega_{\rm S}]$ and
$\displaystyle \Psi_{\rm LIR}=2[(m+1)\lambda_{\rm S} - m\lambda -\Omega]$. 
When Janus is in the inner position, the nearest resonance to the ring edge is the CIR due to Epimetheus located at a semi-major axis of 136684 km. When Janus is in the outer position, the nearest one is the CIR due to Janus at 136660 km.

\section{Averaged equations of motion}
\label{theory_part}

Near a CER, the average motion of a test particle can be described by the following two variables :\\

$\displaystyle \xi = \frac{a-a_{\rm C}}{a} $\\

$ \displaystyle  \psi_{\rm C} = (m+1) \lambda_{\rm S}- m\lambda - \varpi_{\rm S}$,\\

\noindent where 
$a_{\rm C}$ is the CER radius, $\xi$ is the nondimensional distance from the exact CER, 
$m$ is a positive (resp. negative) integer when the particle's orbit is inside (resp. outside) that of the satellite, and $\psi_{\rm C}$ is the CER critical angle.  

The motion near a LER is depicted by the eccentricity vector $(h,k)$ of the particle : \\

$ \displaystyle  h= e \cos(\psi_{\rm L})$\\

$ \displaystyle  k= e \sin(\psi_{\rm L})$, \\

\noindent where the LER argument $\displaystyle \psi_{\rm L}=(m+1) \lambda_{\rm S}- m\lambda - \varpi$
measures the difference between the longitude of conjunction with the satellite and the pericentre longitude of the particle. 

The method to compute the rates of change of $\xi$, $\psi_{\rm C}$, $h$, 
$k$ is detailed in \citet{Araujo17}, based on previous work by, e.g.,  \citet{Shu1984}, \citet{Sicardy-1991} and \citet{Foryta.Sicardy-1996}. The classical technique is to expand the perturbing potential near the first order $m+1:m$ resonance, taking into account the secular terms due to the central planet's oblateness and ignoring the
rapidly varying angles, and then to derive the time variation of the orbital elements from the total averaged Hamiltonian. 
The resulting equations of motion can be written :

\begin{equation}
    \left\{
	\begin{array}{l}
    \displaystyle \dot{\xi} =  - 2 m n \frac{M_{\rm S}}{M} \frac{a}{a_{\rm S}} E e_{{\rm S}} \sin(\psi_{{\rm C}})\\ \\
	\displaystyle \dot{\psi}_{\rm C}    = \frac{3}{2} m n \xi \\ \\
	 \displaystyle  \dot{h} =      - ( \Delta n ) k \\ \\
	 \displaystyle  \dot{k}     = +  ( \Delta n )  h + \frac{a}{a_{\rm S}} \frac{M_{\rm S}}{M}  n A  	
	\end{array}
    \right.  
\label{averag_equ}
\end{equation}
The first two equations of the system (\ref{averag_equ}) describe the CER and the last two describe the LER, where $n$ is the mean motion, $M_{\rm S}$ (resp. $M$) is the mass of the perturbing satellite (resp. the planet), and $\displaystyle \Delta n = \dot{\psi}_{\rm L} \simeq
(m+1) n_{\rm S} - mn$ ($\displaystyle \dot{\varpi}<<n$) is the distance in frequency from the exact LER. The coefficients $E$ and $A$ are combinations of Laplace coefficients $b_{\gamma}^{m}$ \citep{Shu1984} : \\

$\displaystyle E =  -\frac{1}{2} \left[ (2m+1) + \alpha D\right]  b_{1/2}^m (\alpha) \simeq - 0.802 m $\\
		
$ \displaystyle  A=  \frac{1}{2} \left[ 2(m+1) + \alpha D\right]  b_{1/2}^{m+1} (\alpha) \simeq 0.802 m$,\\

\noindent where $\displaystyle \alpha =\frac{a}{a_{\rm S}} $, $\displaystyle D=\frac{d}{d\alpha}$ and where the approximations hold for large $|m|$.\\

We apply the model above by numerically integrating the system (\ref{averag_equ})  with $m=6$, Saturn as central body, Janus or Epimetheus as perturbing satellite, and by changing 
semi-major axes for Janus/Epimetheus every $4.2$ years, which alternately traps 
particles in a CER (first two equations  of system (\ref{averag_equ})) or a LER 
(last two equations). 
This allows us to study in a simple and satisfactory way the dynamics of the ring edge, 
as confirmed by comparative full N-body simulations (Section \ref{num_part}).
Note that equations (\ref{averag_equ}) do not contain any coupling between the CER and LER  as discussed in \citet{Moutamid.etal-2014}, since the two types of resonances are sufficiently isolated in the case of the A ring edge (Figure \ref{jaep_res}). \\

The two CER equations of the system (\ref{averag_equ}) reduce to the simple pendulum equation :

\begin{equation}
    \displaystyle \ddot{\psi}_{\rm C} =  - 3 m^2 n^2 \frac{M_{\rm S}}{M} \frac{a}{a_{\rm S}} E e_{\rm S} \sin(\psi_{\rm C}).
    \label{averag_pendul}
\end{equation}
This equation has the following integral of motion :

\begin{equation}
 \displaystyle J= - \frac{3}{8} \xi^2  + \frac{M_{\rm S}}{M} \frac{a}{a_{\rm S}} E e_{{\rm S}} \cos(\psi_{{\rm C}}) \rm{,}
    \label{jacobi_constant}
\end{equation}
from which we derive the full width of the corotation site, 

\begin{equation}
 \displaystyle W_{\rm C}=    \frac{8}{\sqrt{3}} a_{\rm C} \sqrt{\frac{M_{\rm S}}{M} \frac{a}{a_{\rm S}} |E| e_{\rm S}} \rm{.}
    \label{width_cer}
\end{equation}
This yields $W_{\rm C} = 6.8$ and $4.3$ km in the case of Janus and Epimetheus, respectively.   

The equation (\ref{averag_pendul}) can be developed near the centre of the corotation site ($\psi_{\rm C}=180$ deg, $\xi=0$) to obtain the  
CER libration period : 
\begin{equation}
\displaystyle P_{\rm CER} = 
\frac{P_{\rm ORB}}{|m|} \Big{(} {3\frac{M_{\rm S}}{M} \frac{a}{a_{\rm S}}E e_{\rm S}}\Big{)}^{-1/2}.
 \label{per_cer}
\end{equation} 
We find $P_{\rm CER} = 14.5$ years (Janus) and 23.0 years (Epimetheus). 

The effect of the LER is to induce a forced eccentricity on particles given by the fixed points $\dot{h}=\dot{k}=0$ in Eqs. (\ref{averag_equ}) :

\begin{equation}
e_f= - \frac{a}{a_{\rm S}} \frac{M_{\rm S}}{M}  \frac{n}{\Delta n} A.
\label{e_forced}
\end{equation}
In a frame rotating with the satellite, the particle motions are $(m+1)$-lobed patterns called  streamlines. 
The value of $e_f$ decreases as the distance from the exact resonance increases, 
with a shift of $\pi$ for the streamlines on opposite sides of the resonance. 
The width of the LER is determined by the distance from the resonance 
such that the value of the forced eccentricity is large enough to intersect an 
outer streamline with an inner one \citep{Porco.Nicholson-1987,Murray.Dermott-1999}. 
Then a good approximation for the width of the influence of the LER is given by : 
\begin{equation}
\displaystyle W_{\rm LER} = 2.9 a \sqrt{\frac{M_{\rm S}}{M}}, 
\label{width_ler}
\end{equation} 
i.e. $22.9$ and $12.1$ km in the case of Janus and Epimetheus, respectively.

From Eqs. (\ref{averag_equ}), the oscillation period of the eccentricity vector $(h,k)$ is : 
\begin{equation}
\displaystyle P_{\rm LER} = \frac{2\pi}{| \Delta n |}.
 \label{per_ler}
\end{equation} 
For example with $a=136770$ km and when Janus is in the inner position, $P_{\rm LER}= 6.0$ years. Therefore, the CER and LER libration periods are greater than the
four year period of orbital swap between Janus and Epimetheus.

\section{Numerical simulation}
\label{num_part}

Ring edge particle motions are numerically integrated from the model presented in the previous section. 
These simulations are also compared to full N-body integrations.
The full equations of motion are integrated in a Saturn-centred cartesian reference
frame, using Everhart's 15th-order Radau algorithm \citep{Everhart-1985} and  
taking into account the effects of the planet's oblateness up
to and including terms in $J_6$. State vectors are converted into geometric orbital elements using the algorithm given by \citet{Renner.Sicardy-2006}. Indeed, these elements are not contaminated by the short-period terms caused by planetary oblateness, contrary to the classical osculating elements. Except Janus and Epimetheus, the perturbations from the other Saturnian satellites are neglected. 
Here we summarize the results.

\subsection{Individual particle behaviour}

\begin{figure*}
\centering
  \includegraphics[width=.49\textwidth]{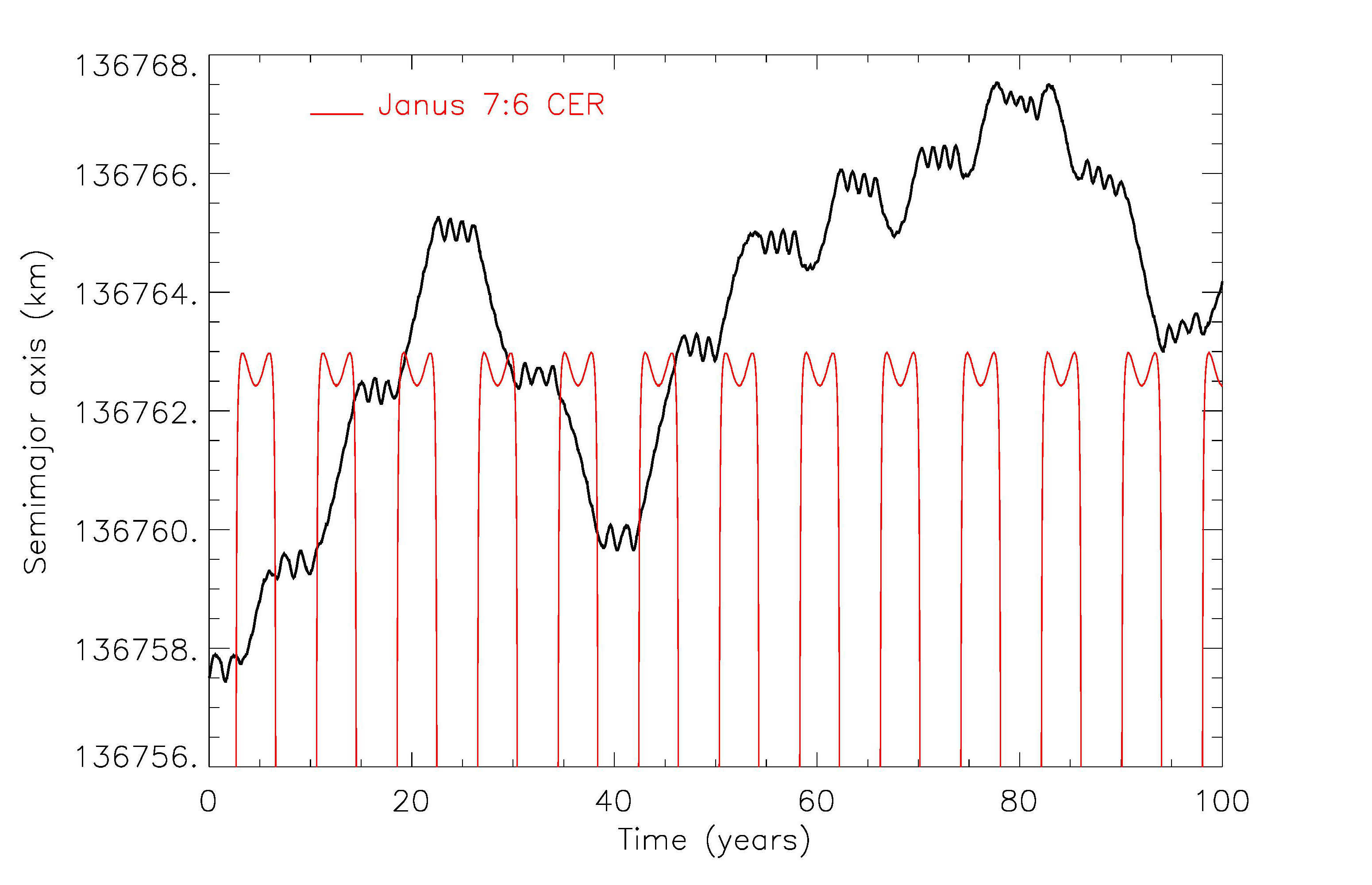}
  \includegraphics[width=.49\textwidth]{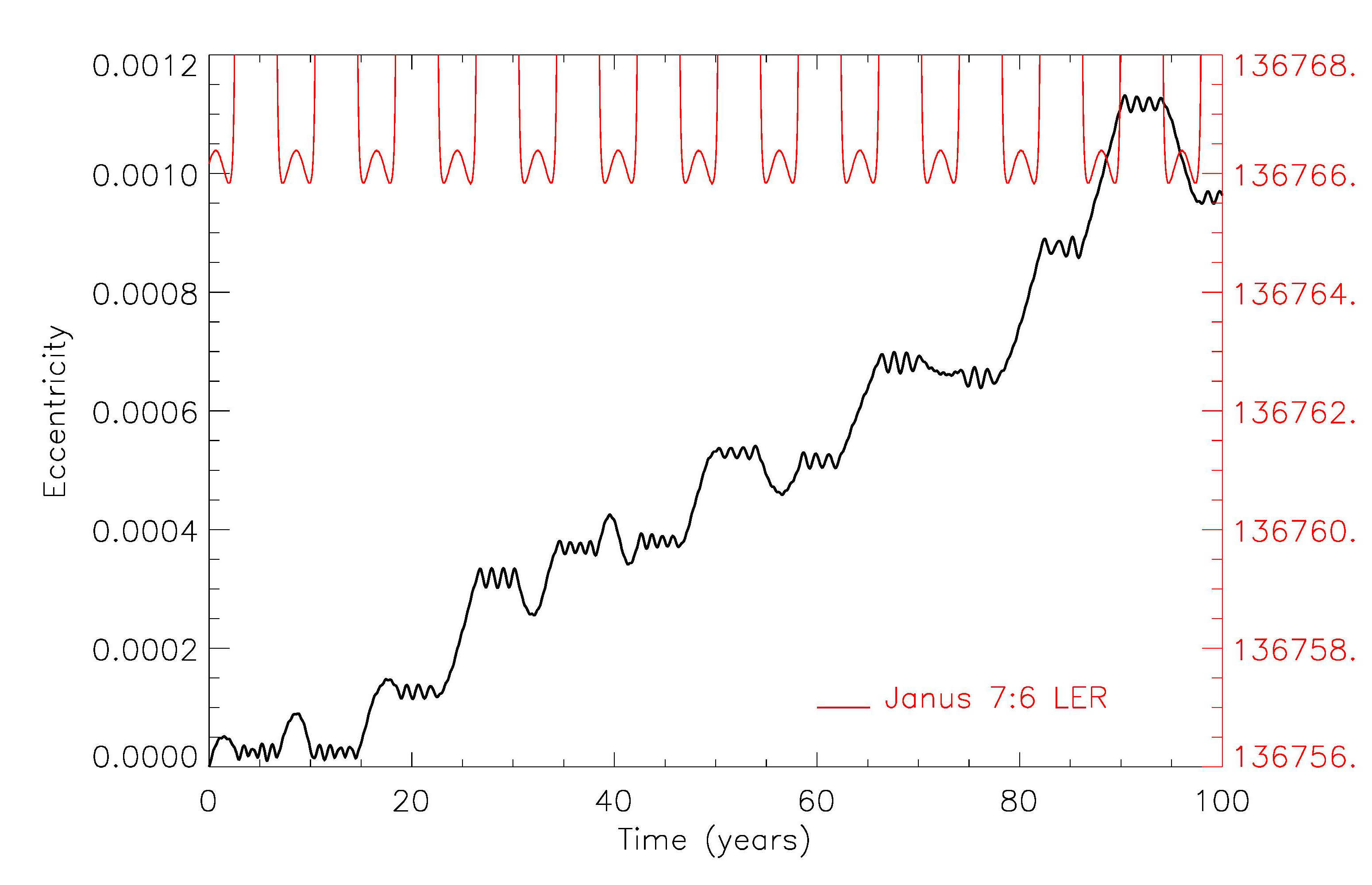}
  \includegraphics[width=.49\textwidth]{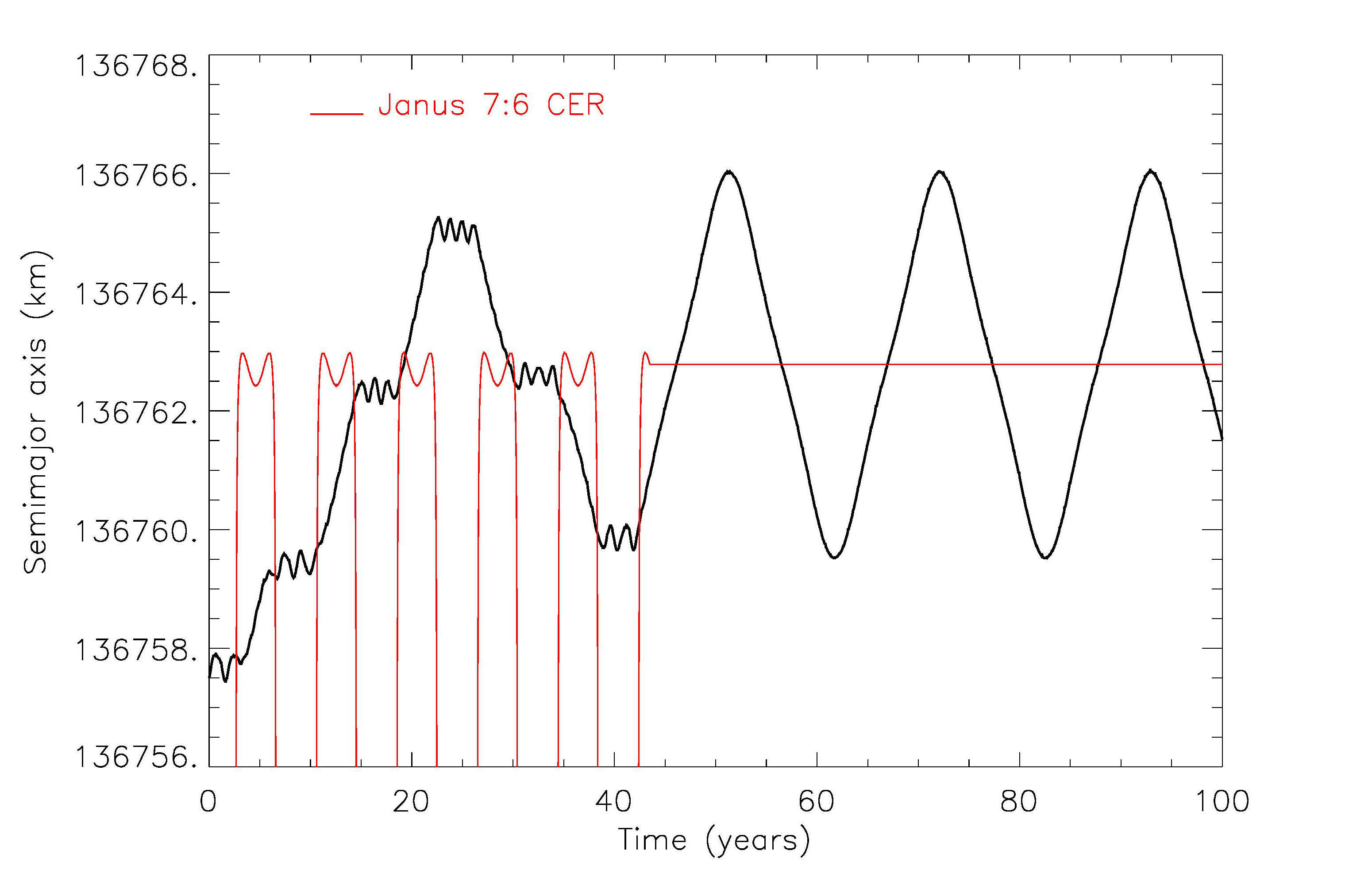}
  \includegraphics[width=.49\textwidth]{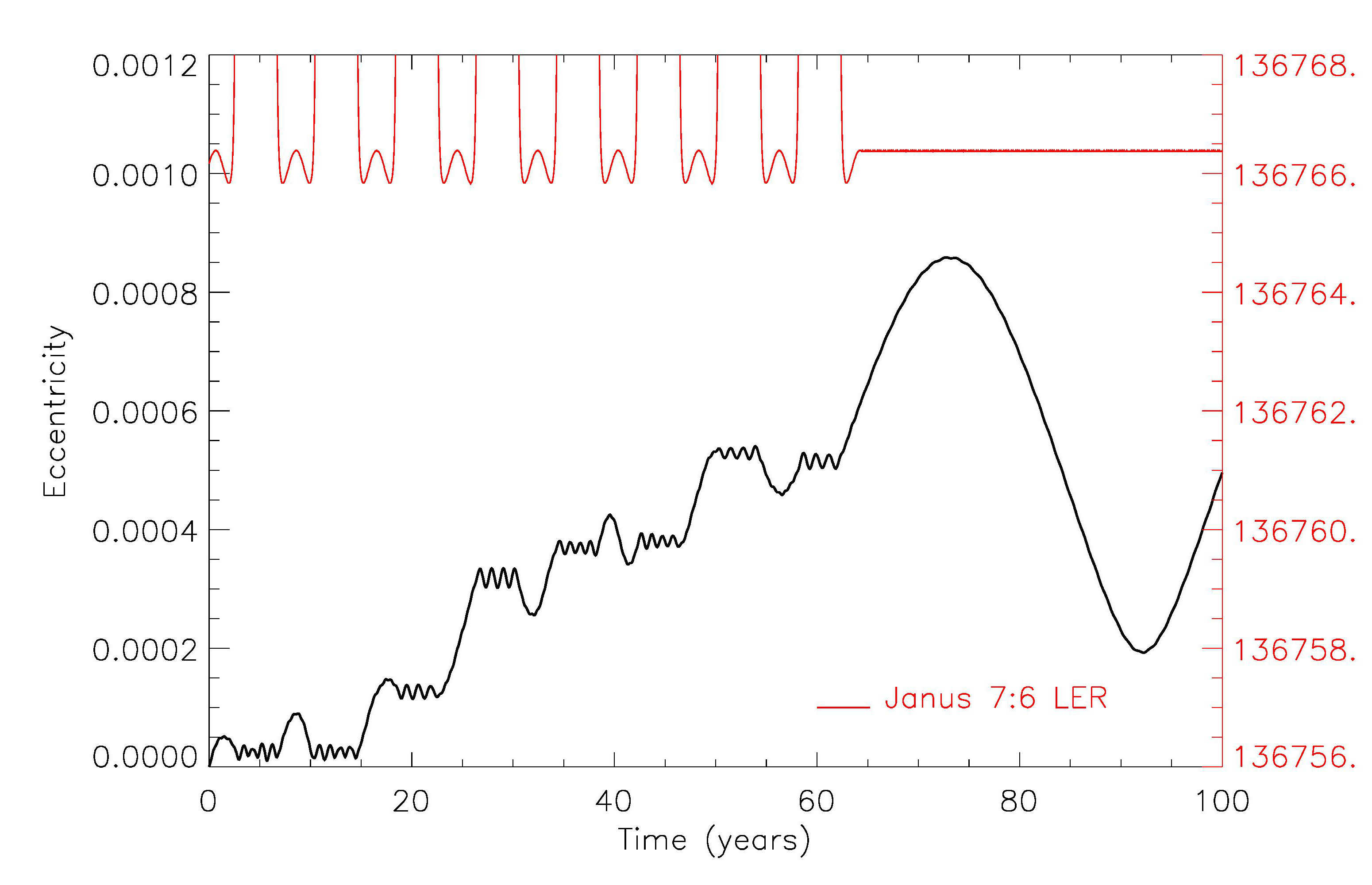}
  \includegraphics[width=.49\textwidth]{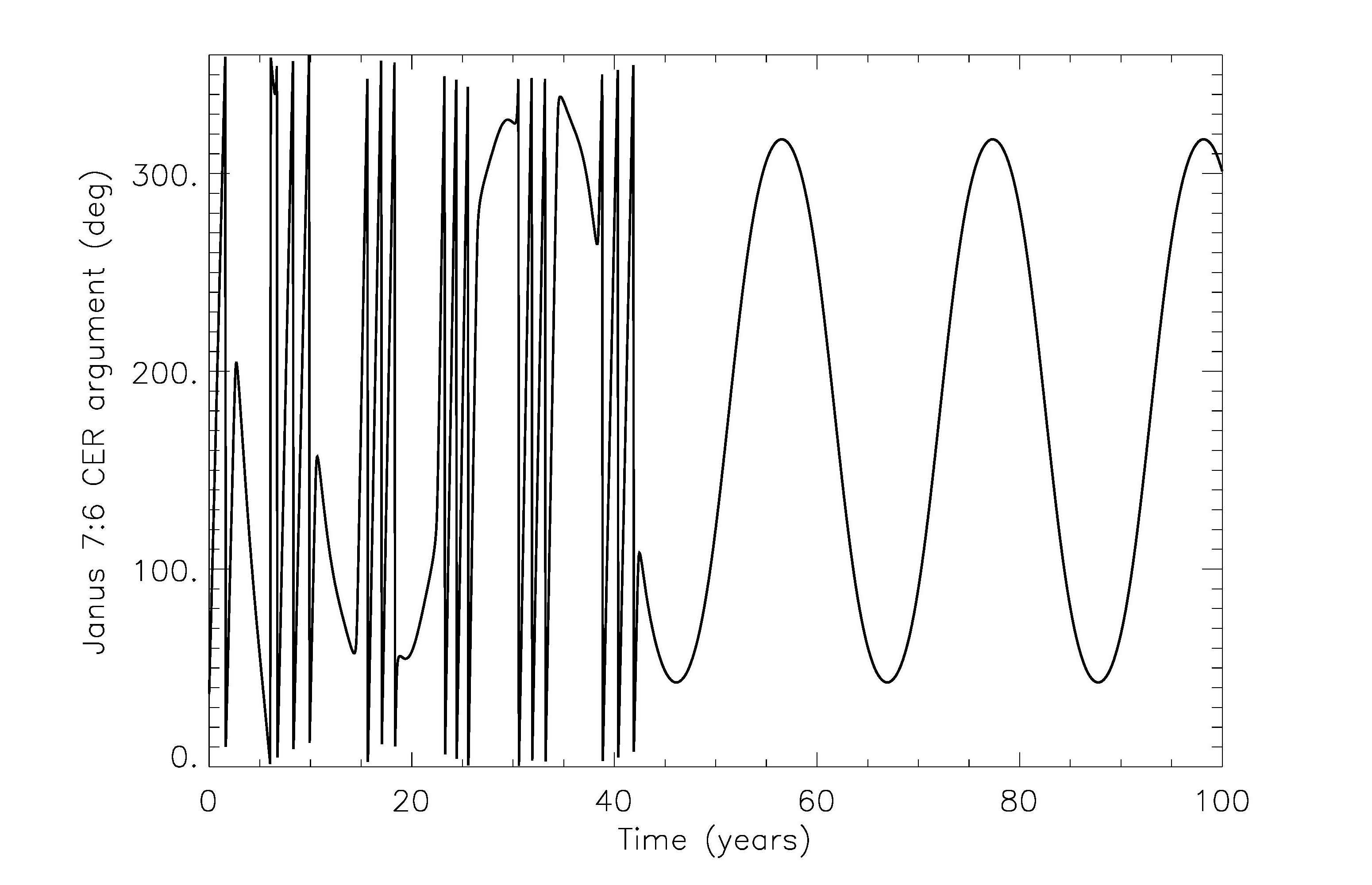}
  \includegraphics[width=.49\textwidth]{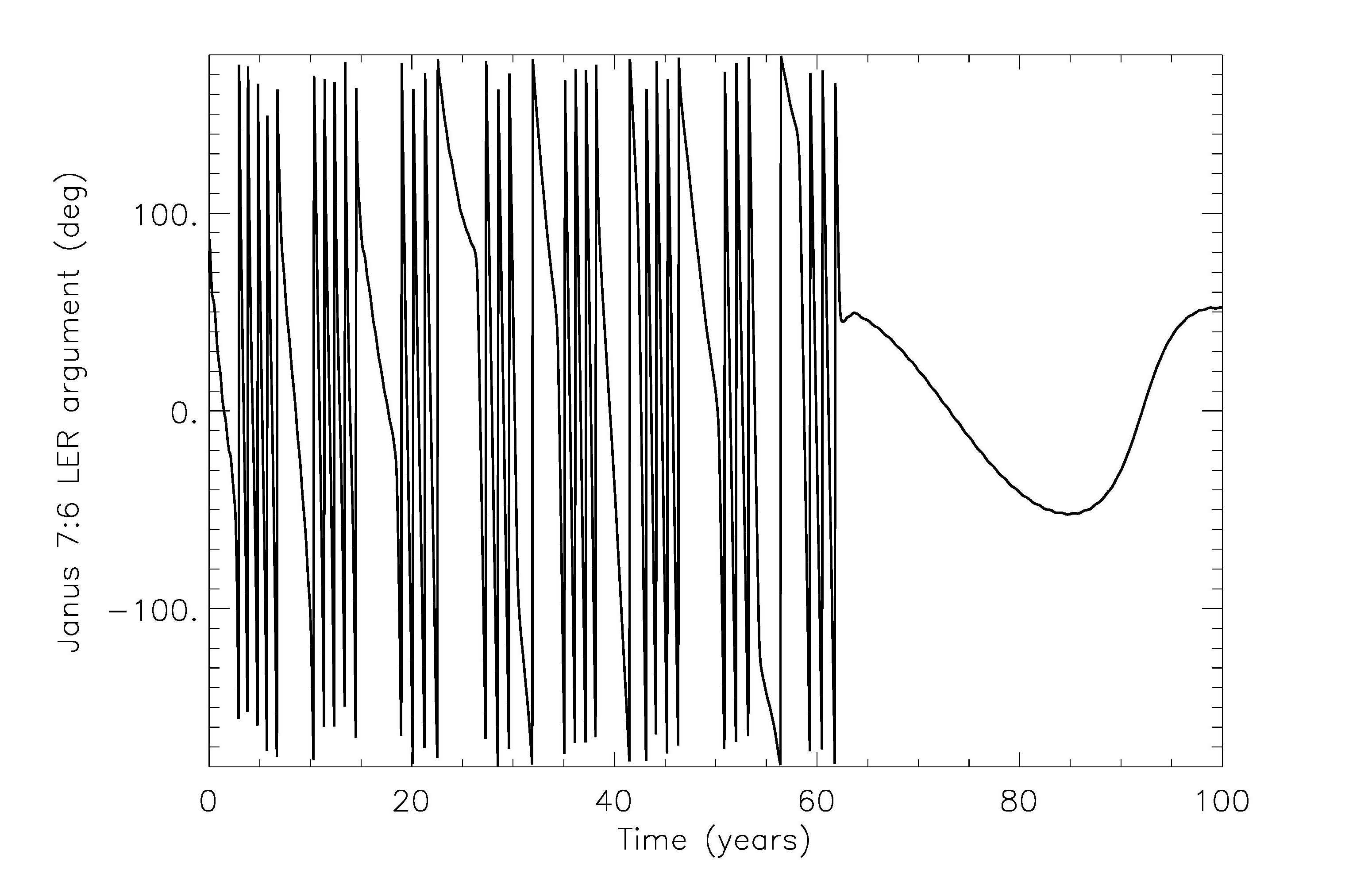}
\caption{Time evolution of a ring edge test particle, alternately trapped in the 7:6 Janus CER and then in the 7:6 Janus LER.   
({\it Top left}) Semi-major axis (black) and location of the 7:6 CER (red). ({\it Top right}) Eccentricity (black) and location of the 7:6 LER (red). ({\it Middle left}) Semi-major axis (black) and location of the 7:6 CER (red), with Epimetheus removed from the simulation at $t=43.5$ years. ({\it Middle right}) Eccentricity, with Epimetheus removed from the simulation at $t=64.3$ years. ({\it Bottom left}) 7:6 CER argument, with Epimetheus removed from the simulation at $t=43.5$ years. ({\it Bottom right}) 7:6 LER argument, with Epimetheus removed from the simulation at $t=64.3$ years.}
\label{indiv_behave}
\end{figure*}

Figure \ref{indiv_behave} typically illustrates the dynamical response of a test particle alternately perturbed by the CER/LER due to Janus-Epimetheus. Initially, the particle has no eccentricity and a semi-major axis $a=136757.5$ km, corresponding to the radial zone perturbed by the Janus resonances only, see Figure \ref{jaep_res}.   

Because of the horseshoe motion, the location of the 7:6 CER shifts from $a=136743.5$ to $136762.0$ km (Table \ref{jaep_res_values}) every $4.2$ years, 
periodically trapping the test particle in this resonance of full width $6.8$ km.
Since the libration period ($14.5$ years near the CER fixed point) is longer than the interval between the orbital switches, the particle does not complete the full libration cycle. 
The particles instead enter corotation sites with random phases, which randomly increase or decrease their semi-major axes as a result of the partial CER libration motion. 
Consequently the motion is chaotic, as shown in the next section (Figure \ref{plot_fli_sma}) : starting with two sets of very close initial conditions can lead to very different final positions after a few passages into resonance. 
Nevertheless, the semi-major-axis range explored by the particle is of the same order as the CER full width.
When Janus is in the inner position, the CER is at 
$a=136743.5$ km well inside the ring particle's orbit, leaving its semi-major axis constant at first order.
In that case the LER location ($a=136766$ km) and the particle's orbit are close, exciting the eccentricity (top right plot of Figure \ref{indiv_behave}). Again, the LER libration period ($6$ years for a particle's semi-major axis $a=136770$ km) is longer than the interval between the orbital switches. After a few trappings in the LER, the eccentricity grows up to $\sim 10^{-3}$, a value in good agreement
with the analytical estimate (equation \ref{ecc_maxeq}).
When Janus is in the outer position, the particle's eccentricity remains almost constant as the LER moves 20 km outwards. 

The middle and bottom plots of Figure \ref{indiv_behave} show that the particle remains trapped into resonance (CER or LER) when we  arbitrarily remove Epimetheus from the simulation at different times, thus cancelling the orbital switches between the two co-orbital moons. 
With Epimetheus removed at $t=43.5$ years corresponding to Janus in the outer position, the 7:6 CER argument librates and the semi-major axis oscillates around the CER radius, 
with an amplitude and a period that compare well with the formulae (\ref{width_cer}) and (\ref{per_cer}), respectively. 
Similarly, the LER argument and the particle's eccentricity
oscillate when Epimetheus is removed at
$t=64.3$ years corresponding to Janus in the inner position, with a period given by equation (\ref{per_ler}).

\subsection{Cumulative effects}

\begin{figure}
\centerline{\includegraphics[width=0.99\columnwidth]{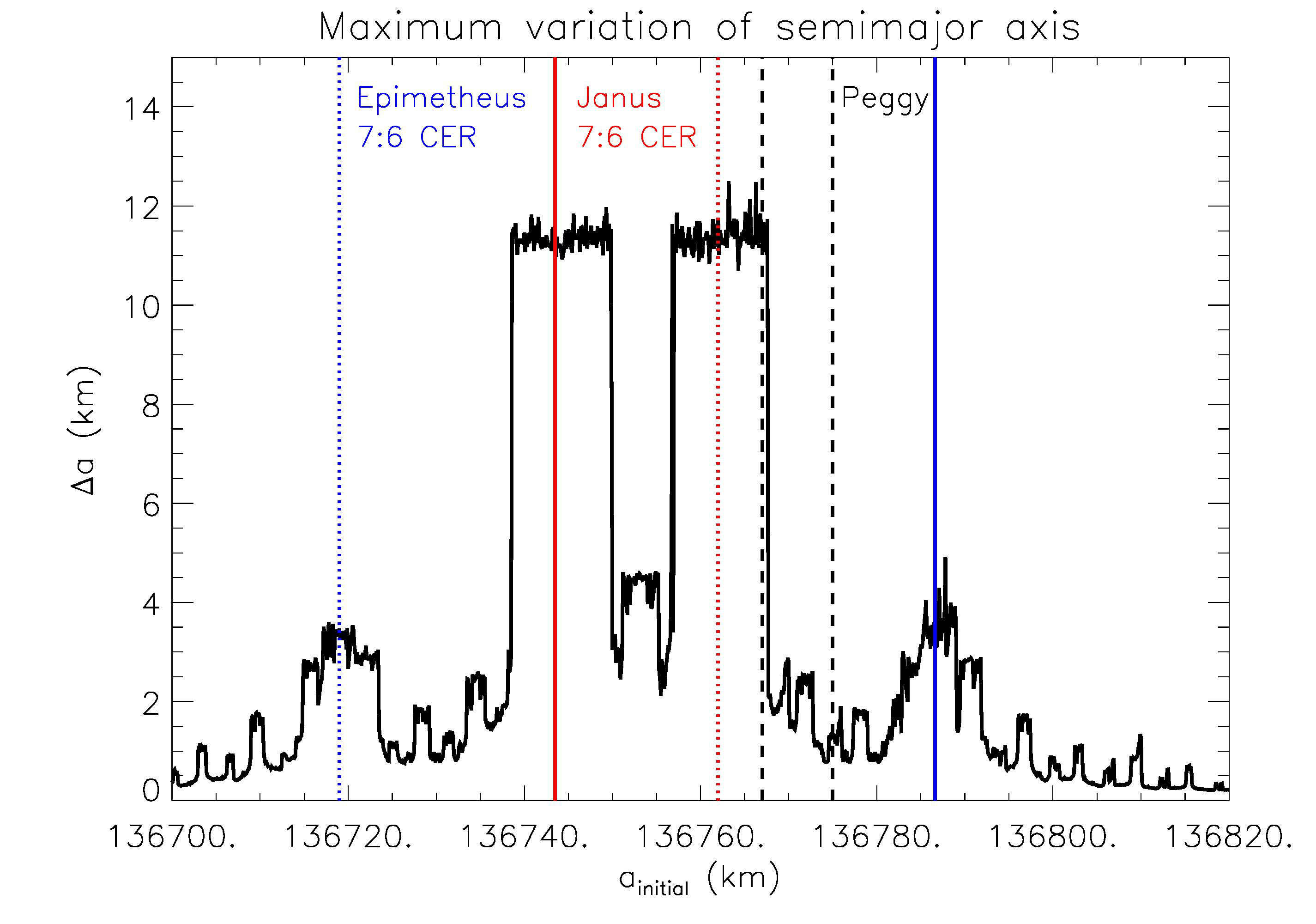}}
\caption{Maximum variation of semi-major axis for ring edge test particles. 
The initial semi-major axes are uniformly distributed between $136700$ and $136820$ km, with a spacing of $0.5$ km. For each given semi-major axis value,
the motion of 10 particles is integrated, varying the initial mean longitude uniformly along the orbit. The particle that experiences the largest variation of semi-major axis is identified. This variation is defined as the difference between the largest and the smallest semi-major axis value reached during the simulation. The integration time is 500 years, corresponding to several dozens of librational paths of Janus-Epimetheus. The locations of the 7:6 CER are indicated in red (Janus) and blue
(Epimetheus). The solid (resp. dotted) lines correspond to Janus in the inner (resp. outer) position. The black dashed lines represents the range of semi-major axes determined for Peggy
(Cooper \& Murray, private communication).}
 \label{plot_delta_a}
\end{figure}

In order to assess the overall effect of the resonances on the ring edge, 
we consider 10000 test particles 
with initial 
semi-major axes uniformly distributed with a spacing of $0.5$ km between 136700 and 136820 km, where the 
 the 7:6 resonances are located (Figure \ref{jaep_res}). 
Initial mean longitudes are also uniformly distributed and eccentricities are fixed at zero. For each given semi-major axis value,
the motion of 10 particles is integrated with an integration time of 500 years, corresponding to several dozens of librational paths of Janus-Epimetheus.
The initial conditions for Janus and Epimetheus are given in Table \ref{tab_orb_elements}. 
We compute the maximum variation of semi-major axis for ring edge test particles, defined as the difference between the largest and the smallest semi-major axis value reached during the simulation. The results are shown in Figure \ref{plot_delta_a}. 
Four radial zones centred on the CER radii (Table \ref{jaep_res_values}) exhibit large semi-major axis variations. These changes in semi-major axis result from the periodic but partial libration motions around the CER fixed points (Figure \ref{indiv_behave} top left). The semi-major axis intervals of the four regions are : 
136715-136723, 136738-136749, 136756-136768, 136783-136791 km. 
The maximum variations reached are respectively $\sim 11$ km and $3.5$ km for particles perturbed by Janus and Epimetheus CERs. The values are larger in the case of the Janus since the latter has a larger mass, and are of the same order as the CER full widths (Table \ref{jaep_res_values}). 
Modulations of semi-major axis for particles located near but outside the resonance (circulation)
and/or the CER sweeping every 4.2 years may explain why the radial zones can be larger than the CER full width, as is the case for Janus CER.\\

\begin{figure}
\centerline{\includegraphics[width=0.99\columnwidth]{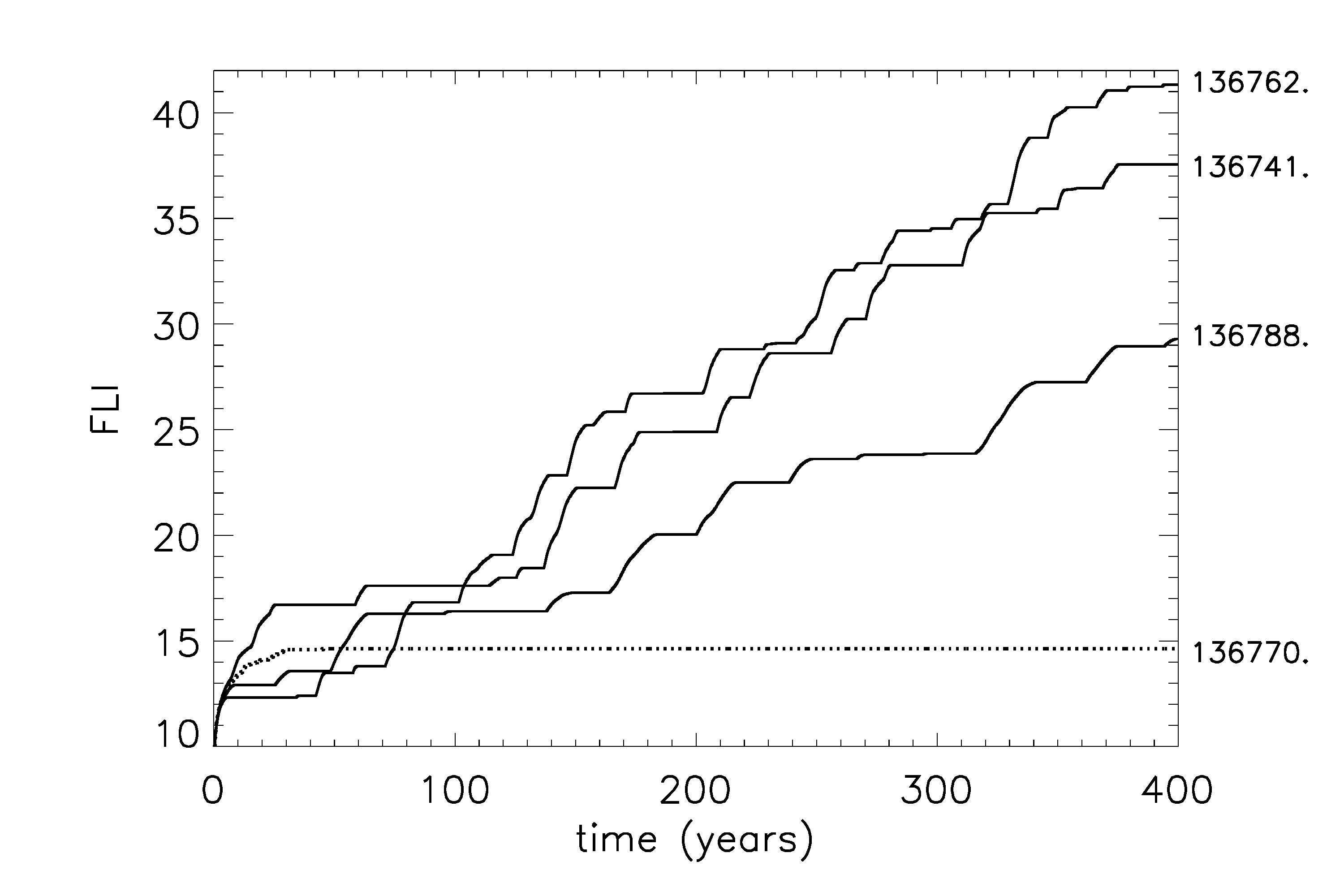}}
\caption{FLI versus time for particles at Saturn's A ring edge. The three upper
curves represent chaotic orbits for particles of the three outer radial zones of Figure \ref{plot_delta_a} (initial semi-major axes 136788, 136741 and 136762 km, respectively). 
The lower dotted curve shows a regular behaviour for a particle at 136770 km, i.e. the typical value defining the ring edge.}
 \label{plot_fli_sma}
\end{figure}

Since particles get into the CERs with random phases, motions in the four radial zones of Figure \ref{plot_delta_a} are chaotic. 
We demonstrate the chaoticity using the Fast Lyapunov Indicator method, hereafter FLI \citep{Fro97,Fro00}. To compute the FLI time evolution, 
the variational equations are integrated simultaneously with the full equations of motion,  
still taking into account the effects of Saturn's 
oblateness (up to $J_6$ included) and using the same physical parameters of Saturn, masses and initial states for Janus/Epimetheus (Table \ref{tab_param_saturn} and \ref{tab_orb_elements}, respectively).
Some details on the method are given in, e.g., 
\cite{Cooper.etal-2015}. 
 For a given autonomous dynamical system $\displaystyle \dot{\bf{x}}=\bf{f}(\bf{x})$, the 
 FLI  is defined  by $\displaystyle \rm{FLI}= \underset{0<t<t_f}{sup} \ln ||\bf{w}(t)|| $, 
 where $\bf{w}$ is a deviation/tangent vector solution of the variational equations 
$\displaystyle \dot{\bf{w}} = \frac{\partial \bf{f}}{\partial \bf{x}}(\bf{x}) \bf{w} $ of the system. 
The FLI is in fact the initial part, up to a stopping time $t_f$, of the computation of the
maximum Lyapunov characteristic exponent, making it a faster tool to distinguish chaotic from regular orbits. 
For a chaotic orbit, the FLI grows linearly with time, whereas the FLI of a regular orbit has a 
logarithmic growth. 
The inverse of the FLI slope defines the Lyapunov time, the characteristic timescale
which measures the time needed for nearby orbits of the system to diverge by $e$. 
Typical results are presented in Figure \ref{plot_fli_sma}. The three upper curves are chaotic orbits 
for particles of the three outer radial zones of Figure \ref{plot_delta_a}, that exhibit large variations of semi-major axes. 
The slopes of the FLI evolution correspond to Lyapunov times of 10-15 years, meaning that just a few CER encounters are sufficient to make neighbouring orbits diverge.  \\

\begin{figure}
\centerline{\includegraphics[width=0.99\columnwidth]{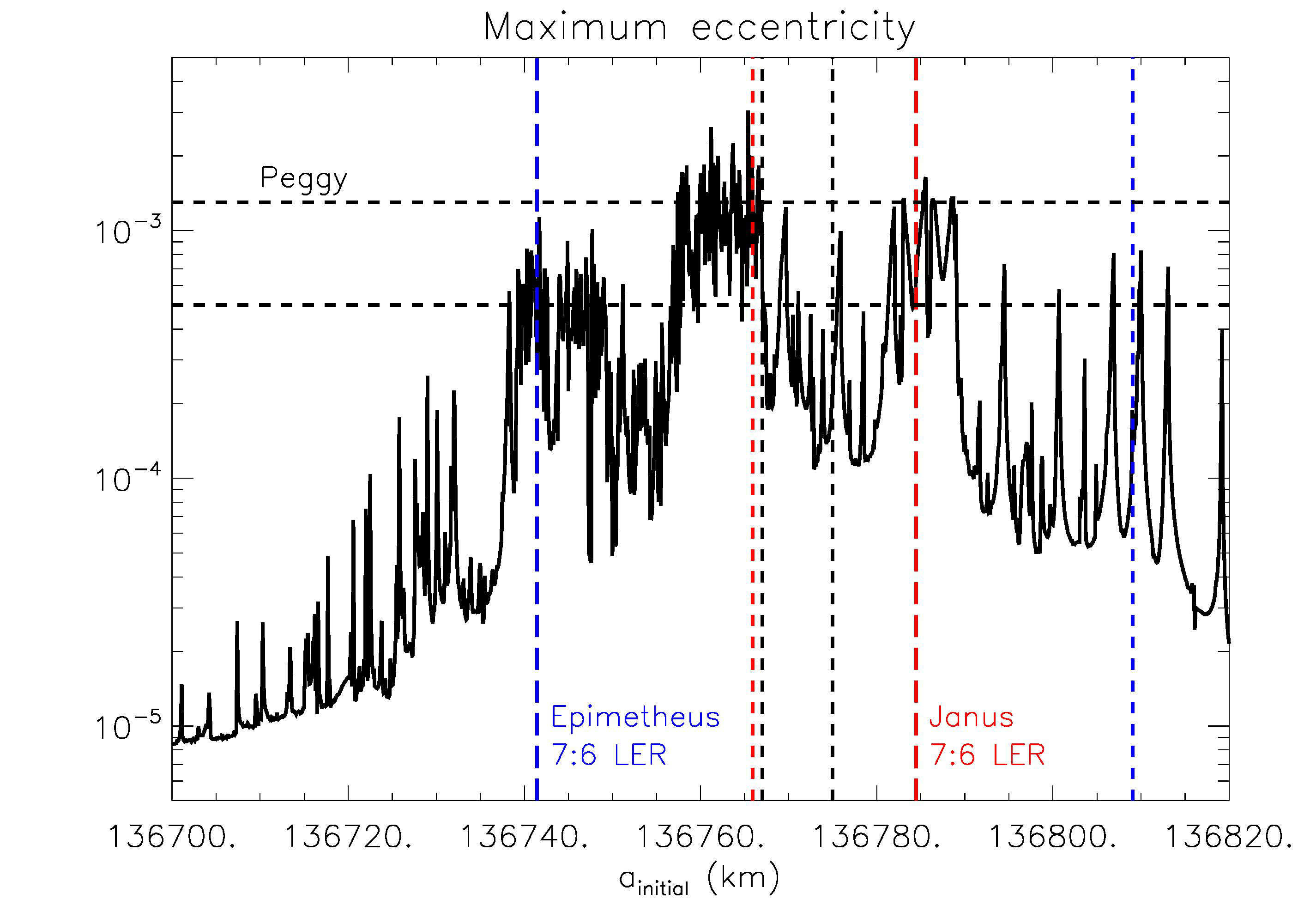}}
\caption{Maximum eccentricity for ring edge test particles. 
The same method as to derive the maximum change in semi-major axis (Figure \ref{plot_delta_a})
is used. The integration time is 500 years. The locations of the 7:6 LER are indicated in red (Janus) and blue
(Epimetheus). The black dashed lines represents the range of semi-major axes and eccentricities determined for Peggy (Cooper \& Murray, private com.).}
 \label{plot_delta_e}
\end{figure}

Figure \ref{plot_delta_e} shows the maximum eccentricity of ring edge particles (initially on circular
orbits), using the same method applied for the maximum changes in semi-major axis. Eccentricities reach values up to $10^{-3}$ for particles at the locations of the 7:6 LER. 
This is consistent with the the analytical expression obtained from classical resonance theory 
for initially circular orbits \citep{Henrard.Lemaitre-1983,Malhotra-1998} :
\begin{equation}
\label{ecc_maxeq}
e_{\rm MAX} = 2 \Big{|} \frac{4}{3} \frac{a}{a_{\rm S}} \frac{A}{m^2} \frac{M_{\rm S}}{M}\Big{|}^{1/3} \rm{,}
\end{equation}
which yields $1.7 \times 10^{-3}$ (resp. $1.1 \times 10^{-3}$) in the case of Janus (resp. Epimetheus). \\

\begin{figure}
\centerline{\includegraphics[width=0.99\columnwidth,angle=0]{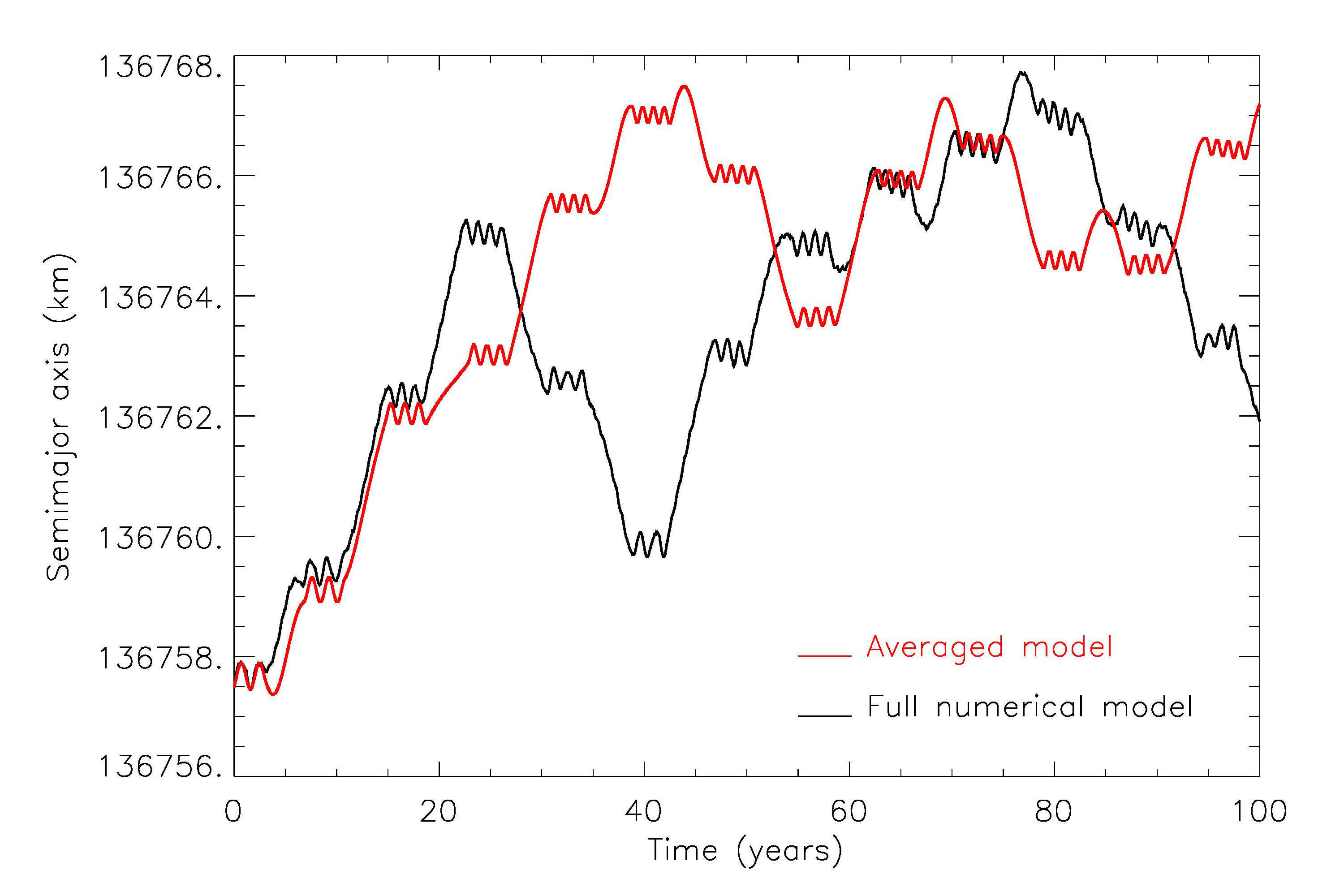}}
\centerline{\includegraphics[width=0.99\columnwidth,angle=0]{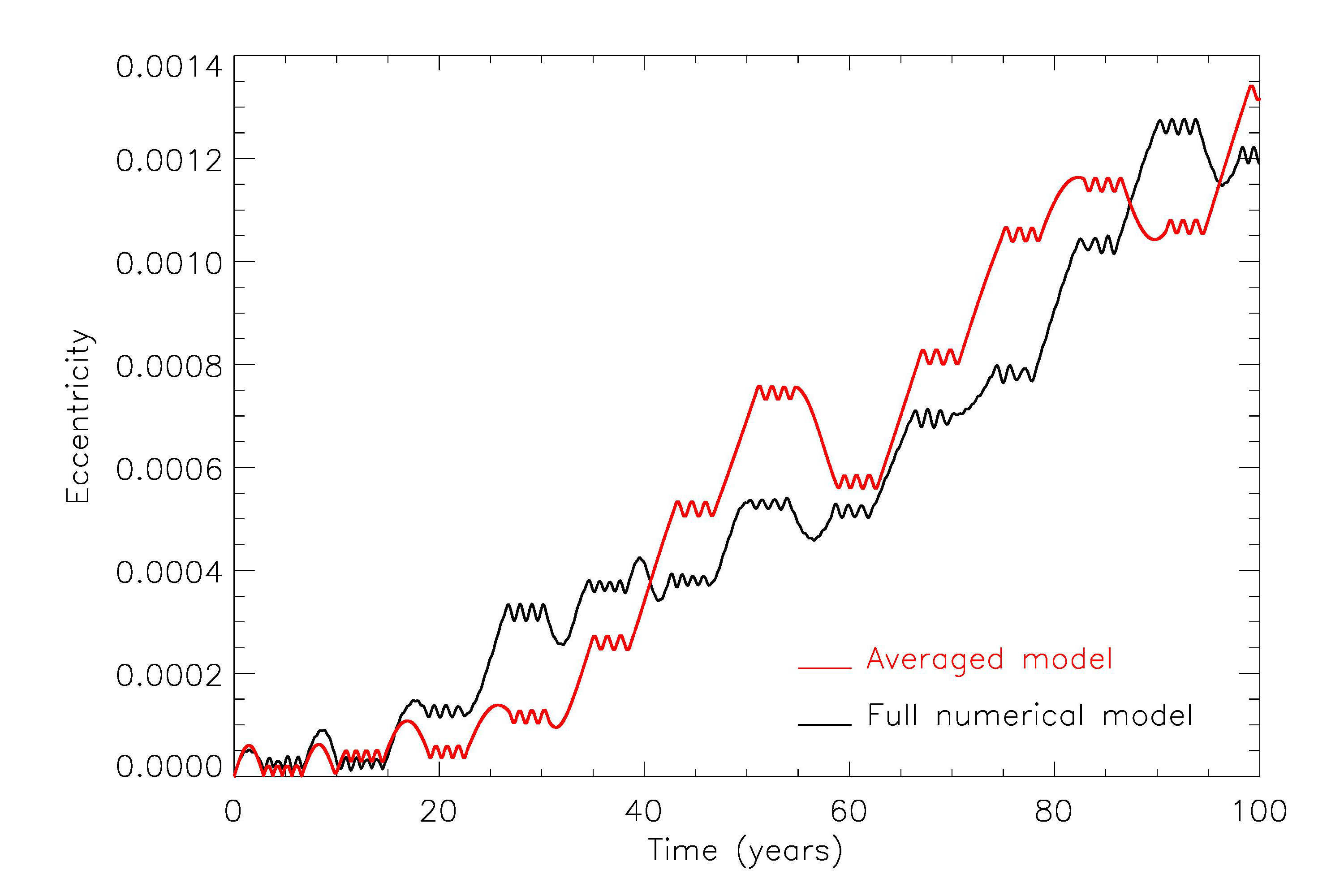}}
\caption{Semi-major axis and eccentricity time evolution for a ring edge particle. 
The red curve is given by the model, and the black curve is from a full
numerical integration including Janus, Epimetheus and Saturn's oblateness up to and
including terms in $J_6$.}
 \label{rk4sma}
\end{figure}

A comparison of the orbital elements
(semi-major axis and eccentricity) as a function of time for ring edge particles is displayed
Figure \ref{rk4sma}. These elements are derived from
both the analytical model (red) and a four-body
simulation (black) including Janus, Epimetheus, Saturn's oblateness 
(up to $J_6$ included), and a test particle. 
The integration of the averaged equations is in very good agreement with the full numerical model, confirming that the system (\ref{averag_equ}) reproduces the main features of the motion. 
Figure \ref{rk4sma} also confirms the short Lyapunov times derived from the FLI computation, 
as very close initial orbits move away after just 2-3 CER encounters.

\begin{figure}
\centerline{\includegraphics[width=0.99\columnwidth,angle=0]{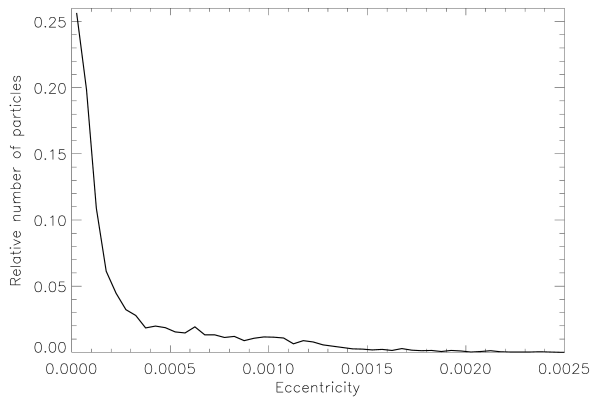}}
\caption{Distribution of particle eccentricities in the ring edge region from a 500 yr simulation. 
The initial semi-major axes are uniformly distributed between $136750$ and $136800$ km.
Particles can reach the theoretical 
maximum value $\sim 10^{-3}$.
Most of the particles have eccentricity values in good agreement with the observed ones 
\citep{Spitale.Porco-2009,ElMoutamid.etal-2016}.}
 \label{histogram1}
\end{figure}

\begin{figure}
\centerline{\includegraphics[width=\columnwidth,angle=0]{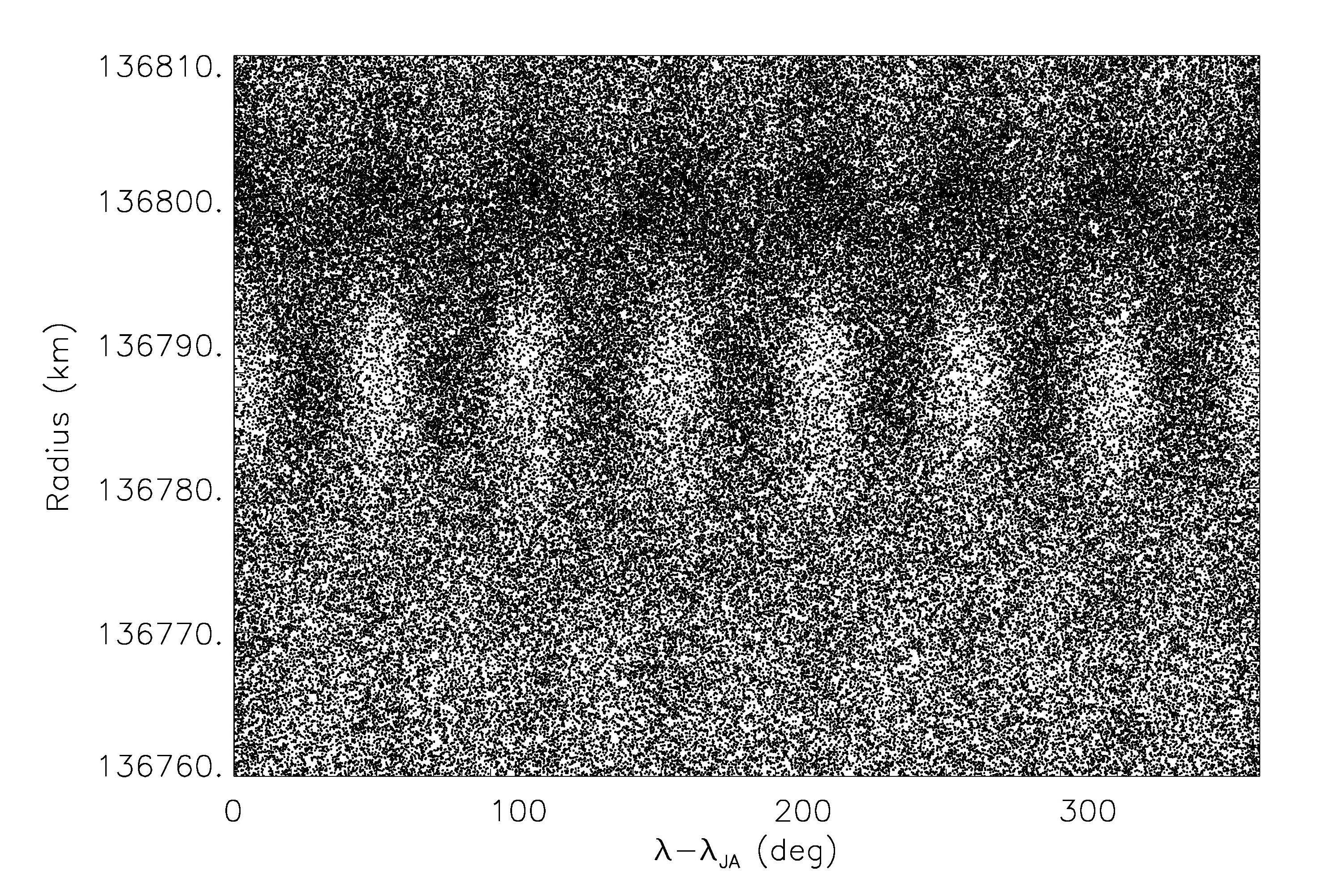}}
\caption{Orbital radii of ring edge particles vs mean longitude in a frame rotating with Janus' mean motion. The plot shows the accumulated positions of particles on 4.2 years, with a time step of 0.1 year and after an integration time of 400 years, when  Janus LER is at 136784 km (Janus in the outer position).}
 \label{accumuJanOuterLindgimp1}
\end{figure}

Particle eccentricities from the model are similar to the observed values derived from Cassini imaging. The histogram Figure \ref{histogram1} essentially shows 
values from $10^{-5}$ to $10^{-4}$, a range comparable to the eccentricity variation among the
 \cite{Spitale.Porco-2009} data sets.
 The $m=7$ pattern resulting from the 7:6 LER is also clearly visible Figure \ref{accumuJanOuterLindgimp1} that shows, as a function of mean longitude in a frame rotating with Janus' mean motion, the orbital radii of ring edge particles accumulated over 4.2 years (after 400 years of integration and with a time step of 0.1 year) when  Janus LER is at 136784 km (outer position). 
The amplitude $ae$ of this pattern is about 15 km, corresponding to an eccentricity of order $10^{-4}$ akin to the value derived from the simultaneous fit to all data sets in \citet{Spitale.Porco-2009}. 
Note that the density of particles 
is lower on a wide orbital radius range between $\sim 136740$ and $136790$ km, since a significant fraction of particles
has reached higher eccentricities due to the 
repeated LER perturbations (Figs. \ref{plot_delta_e} and 
\ref{histogram1}). \\

\subsection{Application to Peggy's orbital elements}
 
The object Peggy is currently not perturbed by the CERs. 
In fact, the semi-major axis range ($136767-136775$ km, 
see Figure \ref{plot_delta_a}) derived for that body from Cassini observations (Cooper \& Murray, private com.) is located right next to the outer chaotic radial zone (centred at 136762 km) due to Janus CER.
On the other hand, the eccentricity interval ($5 \times 10^{-4} - 2 \times 10^{-3}$)
is compatible with the maximum eccentricity forced by the LERs (Figure \ref{plot_delta_e}), 
and matches even better the values around $136760$ km. 
This raises the idea that the object possibly formed in the CER region  
before moving outwards to its current location. 

\begin{figure}
\centerline{\includegraphics[width=\columnwidth,angle=0]{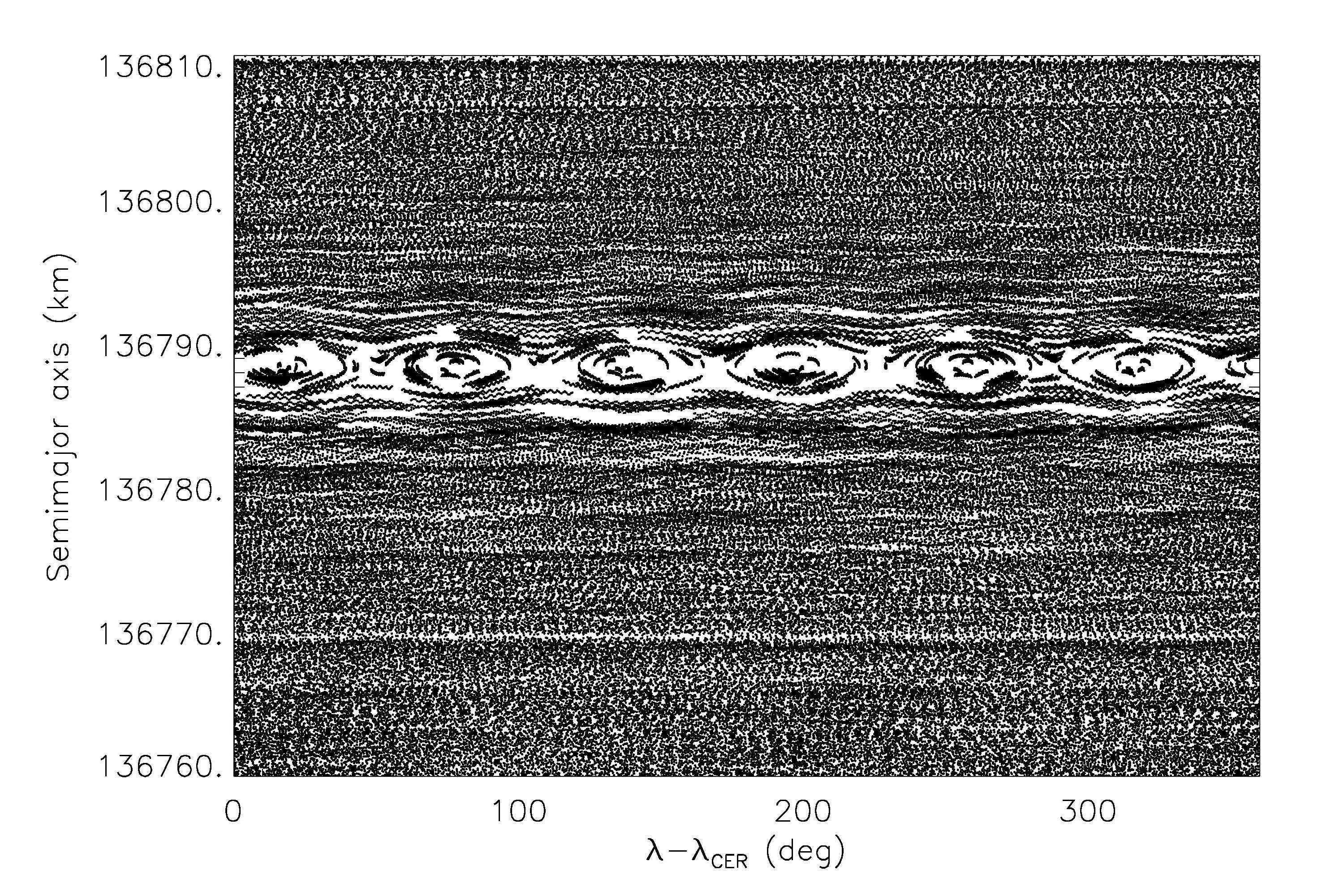}}
\caption{
Semi-major axes of ring edge particles vs mean longitude in a frame rotating at pattern speed $n_{\rm CER}$. 
 The accumulated positions of particles on 4.2 years are displayed,  with a time step of 0.1 year, 
when Epimetheus CER is at 136787 km (Epimetheus  in the outer position). }
 \label{accumuEpiOuter}
\end{figure}

\begin{figure}
\centerline{\includegraphics[width=\columnwidth,angle=0]{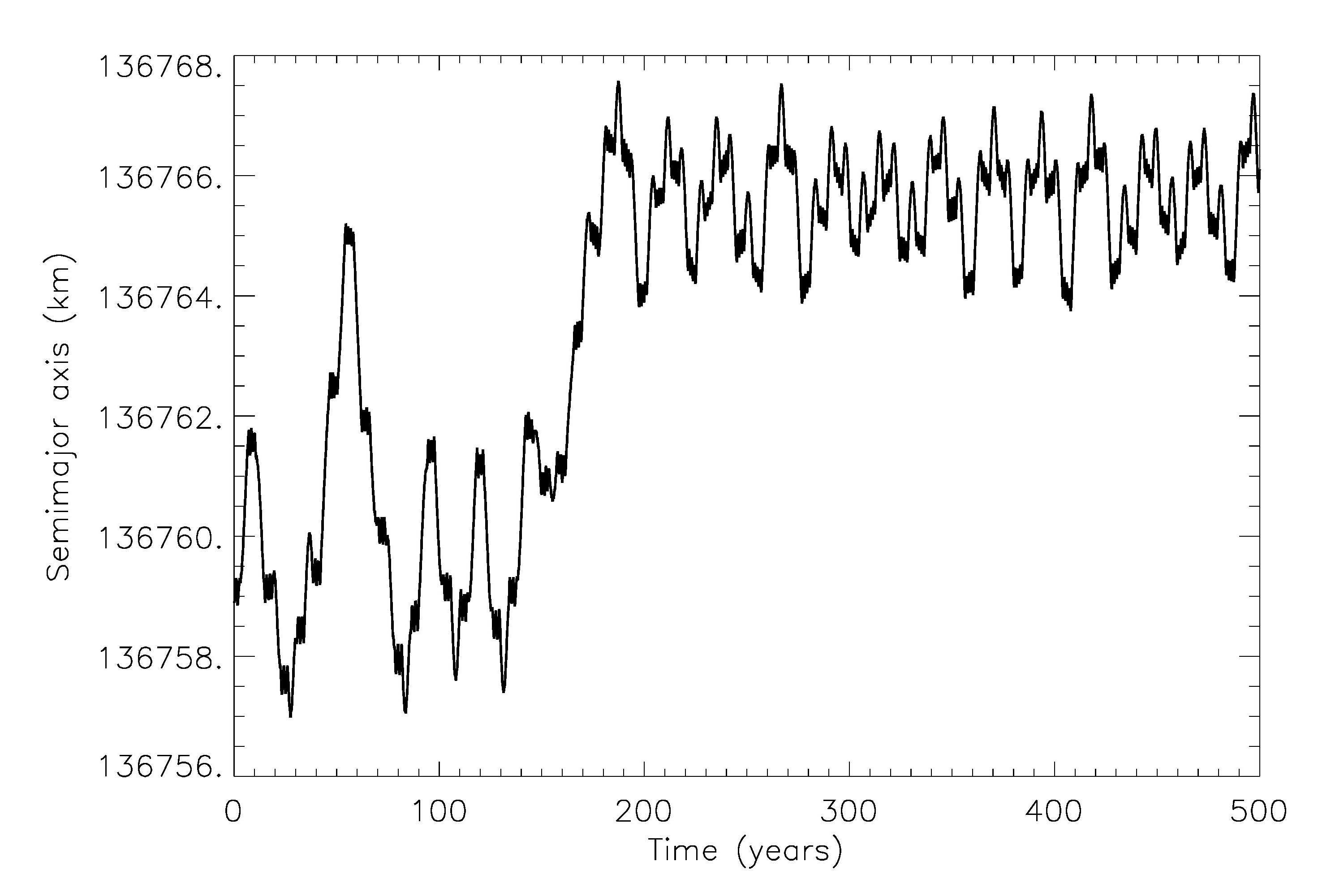}}
\caption{Semi-major axis time evolution for a test particle, driven to the outer part of the 
ring edge by the chaotic crossings of the 7:6 CER with Janus.}
 \label{peggyout}
\end{figure}

In Figure \ref{accumuEpiOuter}, particle semi-major axes are displayed as a function of mean longitude in a frame rotating with the CER pattern speed $n_{\rm CER}$, when Epimetheus CER is located at 136787 km. The six corotation sites are 
distinguishable. The partial libration motions around the CER fixed points create clumpy structures along the ring. These structures, which extend over a few kilometres in the radial direction and have various longitudinal lengths up to dozens of degrees, are undetectable in Cassini ISS data. 
Figure \ref{peggyout} shows an example where a test particle, randomly trapped every 4.2 years  into the 7:6 CER with Janus, moves to the outer part of the ring edge at $\sim 136767$ km, i.e. the lowest possible semi-major axis for Peggy. 
It is therefore possible that the periodic trapping into CERs with Janus/Epimetheus 
may help to aggregate ring particles into larger objects at the ring edge.

\section{Conclusions}

We numerically examined an analytical model that describes the dynamics of the outer edge of Saturn's A ring.
While the seven-lobed radial distortion of the A ring edge resulting from the 7:6 Janus LER is well-known \citep{Porco.etal-1984}, our work is the first study to explain the effects of periodic perturbations 
by CERs, arising from co-orbital satellites in horseshoe motion.  

In terms of orbital elements, we find that the semi-major axis evolution is dominated by chaotic jumps due to periodic captures into CERs with Janus and Epimetheus.
These semi-major axis variations are comparable (or slightly larger) than those expected from the oscillations into the CER sites (Figure \ref{plot_delta_a}). 
Similarly, the eccentricity experiences step-by-step increases 
due to periodic LER perturbations, before eventually reaching the maximum value forced by the resonance (equation  \ref{ecc_maxeq}).
A good agreement is found between the simulated and observed eccentricities of A ring edge particles \citep{Spitale.Porco-2009,ElMoutamid.etal-2016}.

We showed that the object Peggy recently discovered at the ring
edge \citep{Murray.etal-2014}  is strongly perturbed by the Janus 7:6 LER (when Janus is on its inner leg of its libration motion), but not currently perturbed by any CER, with 
fitted semi-major axes just outside one of the four chaotic radial zones (Figure \ref{plot_delta_a}). 
The model presented suggests that the periodic CER/LER perturbations due to the co-orbital moons may help to form objects at the ring edge from particles that move outwards. 
This dynamical mechanism would add to (long-term) satellite formation processes such as the viscous spreading of the rings \citep{Charnoz.etal-2010} or the ejection from the rings of objects denser than ice \citep{Charnoz.etal-2011}.

\section*{Acknowledgements}
N.C.S.A. thanks the Coordena\c{c}\~ao de Aperfei\c{c}oamento de Pessoal de N\'ivel Superior - Brasil (CAPES) - Finance Code 001. 
N.J.C. and C.D.M. thank the Science and Technology Facilities Council (Grant No. ST/P000622/1) for financial support. 
Part of the research leading to these results has
received funding from the European Research Council under the
European Community's H2020 (2014-2020/ERC Grant Agreement no. 669416 ``LUCKY STAR"). 
The authors also thank Thomas Rimlinger for useful comments, the Encelade working group for
interesting discussions, the International Space Science Institute (ISSI) for support, and the members and associates of the $Cassini$ ISS team.





\bibliographystyle{mnras}
\bibliography{references2}


\bsp	
\label{lastpage}
\end{document}